\documentclass[journal]{IEEEtran}
\usepackage{ifpdf}
\usepackage{cite}
\usepackage{amsmath,amssymb,amsfonts}

\usepackage{algorithmicx}
\usepackage{algorithm}
\usepackage{algpseudocode}
\usepackage{array}
\usepackage{url}
\usepackage{color}
\usepackage{xcolor}
\usepackage{graphicx}
\usepackage[caption=false,font=normalsize,labelfont=sf,textfont=sf]{subfig}
\usepackage{float} 
\usepackage{multirow}
\usepackage{enumerate}
\usepackage{diagbox}
\usepackage{bm}
\usepackage{stfloats}
\usepackage{booktabs}

\usepackage{textcomp}
\usepackage{stfloats}
\usepackage{balance}

\begin{document}

\title{Incipient Fault Detection in Power Distribution System: A Time-Frequency Embedded Deep Learning Based Approach
}

\author{Qiyue Li,
\IEEEmembership{Senior Member, IEEE}, Huan Luo, Hong Cheng, Yuxing Deng, Wei Sun, \IEEEmembership{Senior Member, IEEE}, Weitao Li, Zhi Liu,
\IEEEmembership{Senior Member, IEEE}

\thanks{Qiyue Li, Huan Luo, Hong Cheng, Yuxing Deng, Wei Sun, and Weitao Li are with the school of Electrical Engineering and Automation, Hefei Unviersity of Techonology, Hefei, China, (e-mail: liqiyue@mail.ustc.edu.cn; luohuan@mail.hfut.edu.cn; chenghong@mail.hfut.edu.cn; yxdeng@mail.hfut.edu.cn; wsun@hfut.edu.cn; wtli@hfut.edu.cn).}
\thanks{Zhi Liu is with Department of Computer and Network Engineering, The University of Electro-Communications, Japan (e-mail: liuzhi@uec.ac.jp).}
\thanks{Corresponding author: Zhi Liu.}
}

\maketitle
\begin{abstract}
Incipient fault detection in power distribution systems is crucial to improve the reliability of the grid. However, the non-stationary nature and the inadequacy of the training dataset due to the self-recovery of the incipient fault signal, make the incipient fault detection in power distribution systems a great challenge.
In this paper, we focus on incipient fault detection in power distribution systems and address the above challenges.
In particular, we propose an ADaptive Time-Frequency Memory (AD-TFM) cell by embedding wavelet transform into the Long Short-Term Memory (LSTM), to extract features in time and frequency domain from the non-stationary incipient fault signals.
We make scale parameters and translation parameters of wavelet transform learnable to adapt to the dynamic input signals.
Based on the stacked AD-TFM cells, we design a recurrent neural network with ATtention mechanism, named AD-TFM-AT model, to detect incipient fault with multi-resolution and multi-dimension analysis. In addition, we propose two data augmentation methods, namely phase switching and temporal sliding, to effectively enlarge the training datasets. Experimental results on two open datasets show that our proposed AD-TFM-AT model and data augmentation methods achieve state-of-the-art (SOTA) performance of incipient fault detection in power distribution system.
We also disclose one used dataset logged at State Grid Corporation of China to facilitate future research.
\end{abstract}


\IEEEpeerreviewmaketitle

\section{Introduction}

\IEEEPARstart{P}{ower} distribution system delivers electricity from the transmission system to the individual consumers and is an inseparable part of people's lives and society.
Real-time fault detection system plays an important role in maintaining the stability of power equipment \cite{li2022autonomous,li2021optimal,li2022resource}.
In particular, some pre-existing anomalies occur before a fault occurs in the power distribution system, which is called incipient faults \cite{7587342}.

The incipient fault may occur at any time or in any place of the distribution network. Containing a large number of non-fundamental transient signals, the voltage and current time series data shows strong randomness and non-stationary characteristics when incipient fault occurs \cite{SAMET2021107309}. In addition, as the incipient fault of the distribution network is self-recovering and can be self-concealed, only a small amount of data can be logged by traditional fault recorder, which makes incipient faults detection a huge challenge\cite{5590158,8071426,li2022cjpes}.

Detection of incipient fault allows maintenance personnel to replace defective equipment in advance, effectively improving power supply reliability. In addition, it is a kind of predictive maintenance, where the detection of failures at an early stage helps to avoid unexpected disruptions.
There are two mainstream methodologies of fault detection in power distribution system. The first one is traditional fault classification method, in which manually selected features are extracted from the filtered current and voltage time series signal, and then matched with pre-set feature thresholds or patterns to detect the corresponding fault types \cite{5424106,111111,SAMET2021107303,8468069,8626503}. For example, in \cite{9094224}, a method based on human-level concept learning is proposed by selecting waveform features of current and voltage and decomposing it into primitives to detect faults. In \cite{9122438}, an online model based on sequential Bayesian approach is proposed by splitting power quality abnormalities of continuous current. This kind of methods are easy to implement, however, human selected features and thresholds rely heavily on expertise knowledge, and are not sufficiently capable of characterizing complex non-stationary signals to be well applied to incipient fault detection in distribution system.

With the help of artificial intelligence (AI), data driven methods are also applied to incipient fault detection in power distribution system \cite{8968751,8370789,8354947,9702755}. Due to the complex causes and electrical characteristics of incipient faults in power distribution system, it is difficult to establish a comprehensive mathematical model. On the other hand, data driven methods are more effective to deal with incipient fault diagnosis in the power distribution network, and can detect some unknown faults. For example, Long Short-Term Memory (LSTM) utilizes memory units instead of hidden layers in traditional Recurrent Neural Network (RNN), which constructs a more powerful model over time series using contextual information, and shows good performance on time series estimation. In particular, regarding the voltage and current time series data, LSTM cell is utilized to build a deep RNN architecture to automatically extract features and perform fault detection \cite{9177071}. In \cite{SKYDT2021108691}, a LSTM based network is proposed by learning low-resolution data from a real case study to detect incipient faults.

However, due to lack of the frequency domain analysis, LSTM based schemes cannot fully extract features of time series data, especially for the non-stationary incipient fault signal \cite{LIU2021113917}. Besides, incipient faults in power distribution system are usually of short duration and self-recovering,
which leads to the unavailability of sufficient samples to train the LSTM network. And thus exacerbates the difficulty of data-driven incipient fault detection methods.

In this paper, to improve the feature extraction ability from random and non-stationary signal of incipient faults, we propose an ADaptive Time-Frequency Memory (AD-TFM) cell which embeds adaptive wavelet transform into LSTM. 
Specifically, we first use wavelet transform to decompose different frequency signals existing in a non-stationary signal into non-overlapping frequency bands. At each time step, the learned wavelet transform coefficients are multiplied by the input signal to obtain the time and frequency domain features. The coefficients are stored in AD-TFM and propagated to next time step to improve feature extraction abilities. In addition, we make the scale parameters and translation parameters of wavelet transform learnable to automatically adapt to the input signal, which can achieve multi-resolution and multi-dimensional analysis of non-stationary incipient fault signals.

Then we construct an AD-TFM cell based RNN model to perform incipient fault detection in power distribution system. 
To focus the neural network on global hidden information, we strengthen the stacked AD-TFM network by adding an ATtention layer, i.e., a new AD-TFM-AT model is implemented.
The correlation of the hidden state output at all time steps of AD-TFM cells are calculated in the attention layer. Then, the correlation degrees are used to make a weighted average of hidden states of all time steps.  
By improving the attention of neural network to the time steps containing fault information and increasing the feature extraction ability of hidden information for all time steps, the fault detection accuracy is further improved.
To enlarge the training sample of incipient fault data and improve detection performance of AD-TFM-AT, we propose two data augmentation methods, namely phase switching and temporal sliding.
Based on the available small incipient fault dataset\cite{bwjy-7e05-19} and a relatively large dataset logged in State Grid Corporation of China, our proposed method achieves state-of-the-art (SOTA) performance.

The main contributions of this paper are as follows:
\begin{enumerate}
\item We propose an AD-TFM cell based on adaptive wavelet transform, which performs feature extraction at different scales to effectively deal with the non-stationary incipient fault signals of power distribution system. We design AD-TFM-AT, an attention assisted AD-TFM based RNN model, which increases the weight of the most relevant hidden states in fault detection and guides the feature fusion process.

\item We propose two effective data augmentation methods, i.e., phase switching and temporal sliding, which swap the phases of the voltage and current data of the fault signal with the remaining phases and 
intercepts each fault with different starting points, respectively.
These methods effectively expand the small fault data set in the power distribution system and improve the training performance.

\item We conduct extensive experiments and experimental results on two open datasets, annd the results show that our proposed AD-TFM-AT model and data augmentation methods achieve SOTA performance of incipient fault detection in power distribution system.
We also disclose a relatively large dataset logged at State Grid Corporation of China to facilitate future research\footnote{https://github.com/smartlab-hfut/SGAH-datasets}.

\end{enumerate}

The rest of the paper is organized as follows.Section \ref{Related Work} presents the latest research related to incipient fault detection.
Section \ref{Analysis of initial fault signals} explores the random and non-stationary characteristic of incipient fault in distribution network based on a simplified circuit model.
Section \ref{Wavelet_based_LSTM_cell} introduces our proposed AD-TFM cell based on adaptive wavelet transform and LSTM. Section \ref{Fault_Detection_Model} shows the hierarchical structure diagram of the AD-TFM-AT, and two methods of data augmentation are explained in Section \ref{Data_augmentation}. In Section \ref{sec_Experiments}, we show the performance of data augmentation and incipient fault detection accuracy against two datasets. Finally, Section \ref{sec_conclusion} concludes this manuscript. 

Note that this paper is an extended version of our previous conference paper \cite{9713456}. Different from \cite{9713456},

this paper analyzes the non-stationary
of fault signals based on a simplified distribution network circuit model. In addition, an adaptive wavelet transform with learnable wavelet scale parameters and translation parameters is proposed, which realizes the multi-resolution analysis of fault signals. We also added an attention mechanism to the neural network to enhance the focus on time step hidden states that are highly correlated with fault classification. This further improves the network's ability to detect incipient faults.

\section{Related Work}\label{Related Work}
This section explains the latest research related to incipient fault detection.

\subsection{Faults in power distribution systems}
The incipient faults are transient events that occur at random locations and are pre-emptive hidden faults before permanent faults occur. 
Incipient faults in the power distribution network usually occur in underground cables\cite{8716299,9160622,SAMET2021107309}, transformer equipment\cite{9166441,LOPES2021107519}, distribution networks with high Distributed Energy Resources (DERS) penetration\cite{9866605}, and so on. 
There are many reasons for incipient faults, such as tree interference, animal contact and coil contact \cite{9094224}. When an incipient fault occurs, the fault phases voltage and current change, waveform is distorted, and the fault transient signal exhibits non-stationary characteristics. Meanwhile, incipient faults are typically self-clearing faults and have a short duration, ranging from a quarter of a cycle (sub-cycle), to up to four cycles (multi-cycle) \cite{9094224}. Thus, incipient faults are less well documented, and detection methods for incipient faults are much needed.

\subsection{Traditional fault detection methods}

Traditional fault detection methods, which mainly include similarity detection\cite{9861731,9870694}, waveform eigenvalue decomposition\cite{9424578,8998388,9416172}, and model parameter estimation\cite{9874258,9381622,9489316}, have been used for fault detection in power distribution system. For example, in\cite{9861731}, an expression for the transient zero sequence current characteristics under the influence of the inverter is derived, and a method based on the first-order accumulated generation operator (AGO) and the improved cosine similarity is proposed to identify the faulty feeders. In\cite{9424578}, the wavelet singular value decomposition is applied to obtain the edge components of the normalized fault current amplitude for fault detection. In\cite{9489316}, the transformer state space model, linear parameter varying (LPV) observer, primary and secondary voltage values are used to estimate the primary current at each time step of the transformer. The estimated primary currents are compared with the actual primary currents to distinguish whether the transformer is internally or externally faulty.

The above methods rely on the manual extraction of features and then achieve fault identification based on rules or thresholds set by manual experience. In distribution networks with different parameters, the threshold values set by various methods vary greatly. Moreover, these methods lack the analysis of the non-stationarity of faults, which limits the application and effectiveness of traditional fault detection methods.

\subsection{AI based fault detection methods}
With the wide application of AI, methods based on machine learning and deep learning are applied to fault detection.
From the computer vision point of view, there are schemes using Convolutional Neural Networks (CNN)\cite{9381699,LI2022108148}. Recently, hybrid approaches for fault recognition have also been proposed \cite{9696000,9089306}.
For example, in \cite{9795900}, zero-sequence currents are transformed into spectrograms by short time Fourier Transform, and then a two-channel CNN is constructed to achieve fault classification. In \cite{9698845}, a Multi-layer Long Short-Term Memory Network (MLSTM) is applied to voltage waveform analysis in order to detect whether a fault occurs in the grid. In \cite{9696000}, a hybrid statistical learning and machine learning approach is proposed to identify fault-inducing regions in photovoltaic (PV) farm based on micro Phasor Measurement Unit (PMU) measurement data. 
However, the above AI-based methods lack the analysis of fault signal features, and do not fully take into account the non-stationary nature of incipient faults in the power distribution network.

\subsection{Solution with insufficient training data}
Fault data scarcity is also an important issue faced by fault detection using deep learning methods, and several methods are proposed to cope with this problem \cite{9775087,9733339}.
In particular, in \cite{9775087}, various pre-trained models are fine-tuned on different substations through migration learning and federated learning. However,\cite{9775087} requires a large number of deployable substation resources and edge-cloud communications, which will incur a large cost overhead.
In \cite{9733339}, the fault current data are decomposed into multilayer wavelet coefficients which are fused into a matrix. Then the matrix is mapped into a phase space image with three channels (RGB) by colormap indexes as the input of the classification model. Among them, two data enhancement methods are proposed. The first one is to change the colormap indexes randomly to obtain phase space images with different color domains. The second one is to convert the phase space images from RGB color mode to HSV (Hue, Saturation, Value) mode. The Hue channel value is changed to generate different images. Then the different images are changed back to RGB mode to achieve data enhancement.
In \cite{9733339}, only the color mapping index and Hue of the phase space images of the coefficient matrix are changed to obtain a different color graph of the same fault data, and no new fault information is actually generated. 
In summary, the existing methods mainly address the problem of insufficient incipient fault data in distribution networks through migration learning and fault image data enhancement. These methods perform data enhancement in terms of increasing the fault data acquisition surface or changing the fault data mapping. And they do not actively generate new fault information nor consider fault characteristics.

\section{Incipient faults and their features} 
\label{Analysis of initial fault signals}

\begin{figure*}[htbp]
\centering

\subfloat[]{
\includegraphics[width=3.4in]{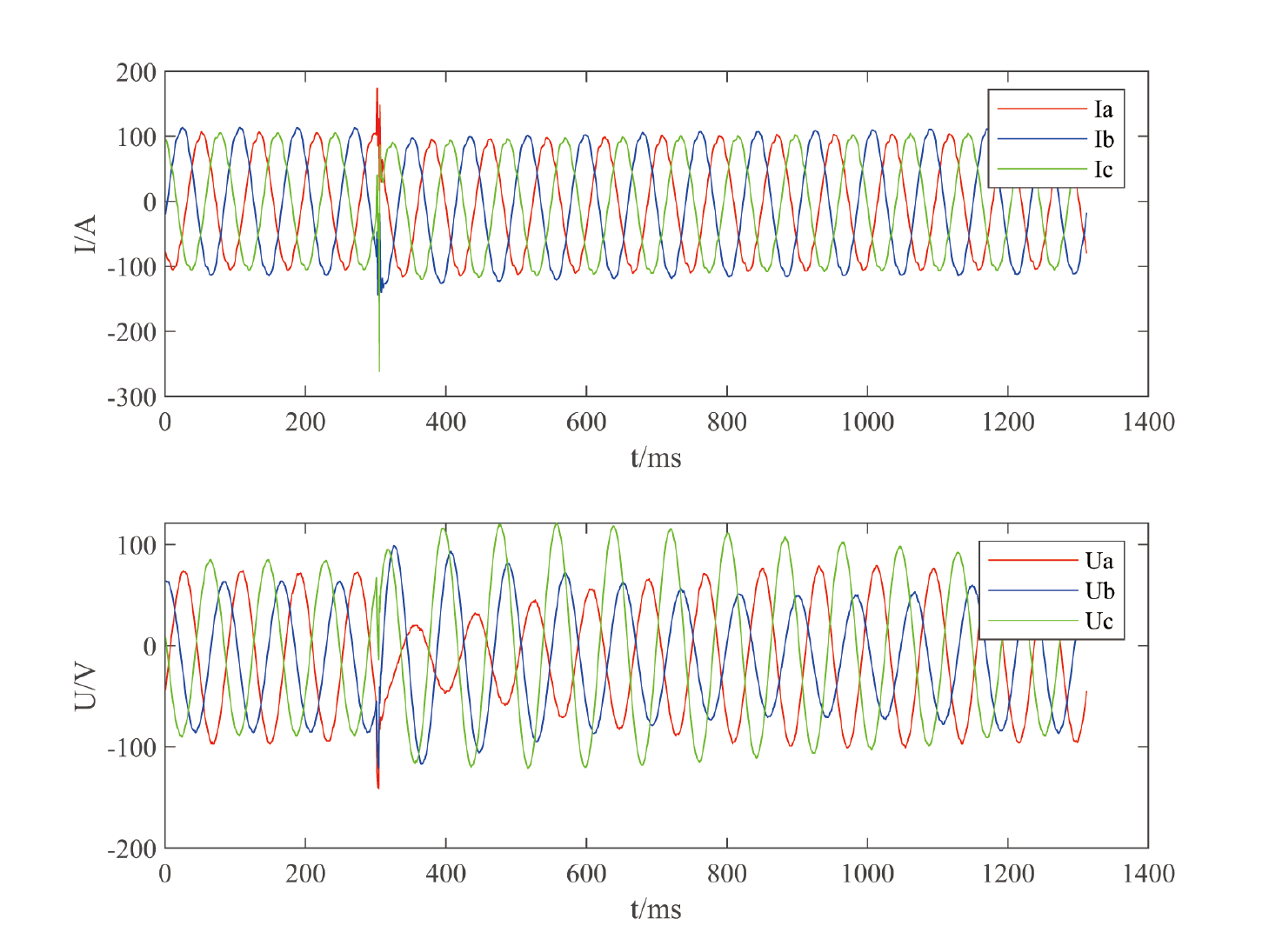}%
\label{fig_subcycle}
}%
\hspace{-8mm}
\subfloat[]{
\includegraphics[width=3.4in]{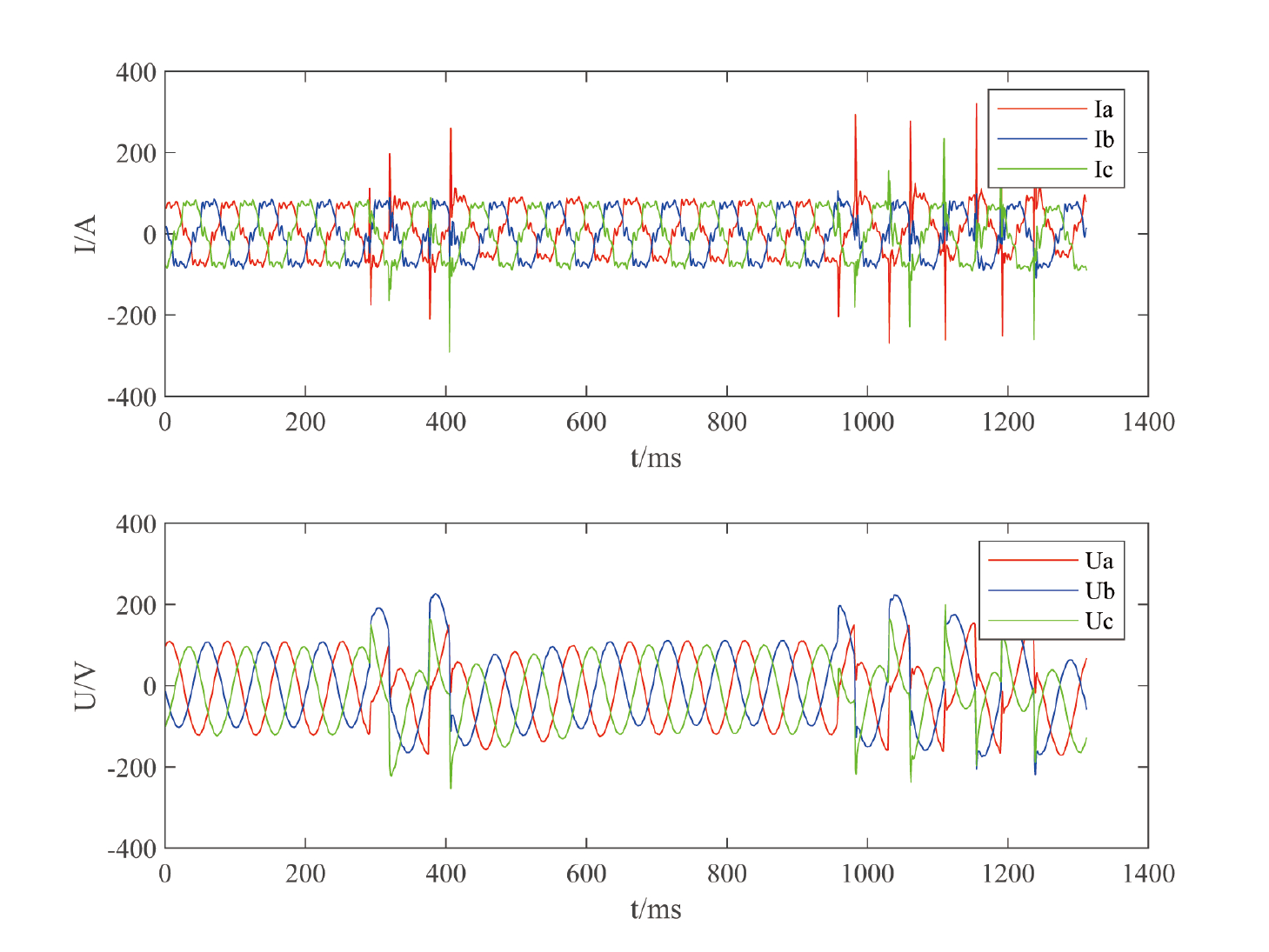}%
\label{fig_multicycle}}

\quad
\subfloat[]{\includegraphics[width=3.4in]{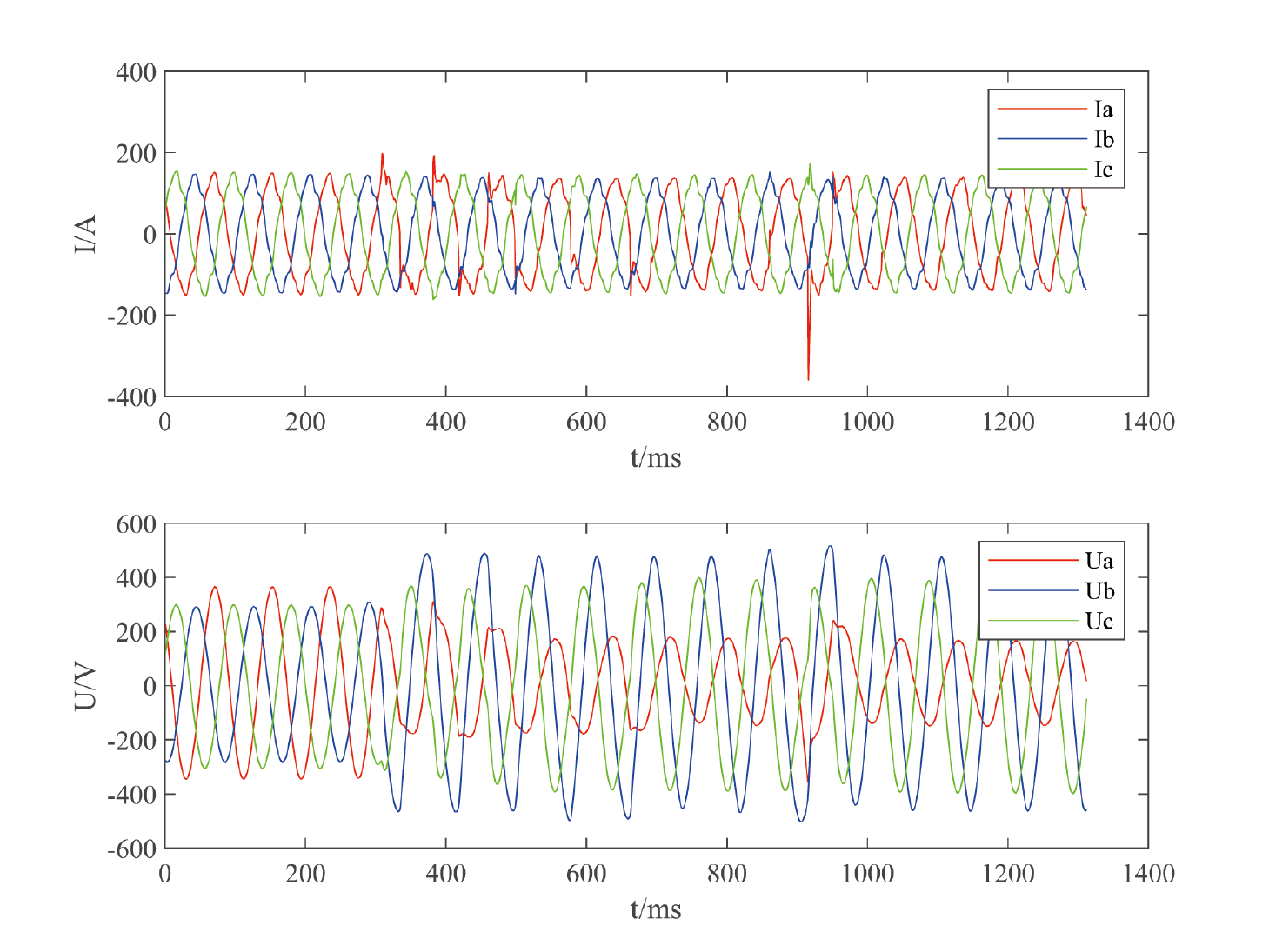}%
\label{fig_high}}
\hspace{-8mm}
\subfloat[]{\includegraphics[width=3.4in]{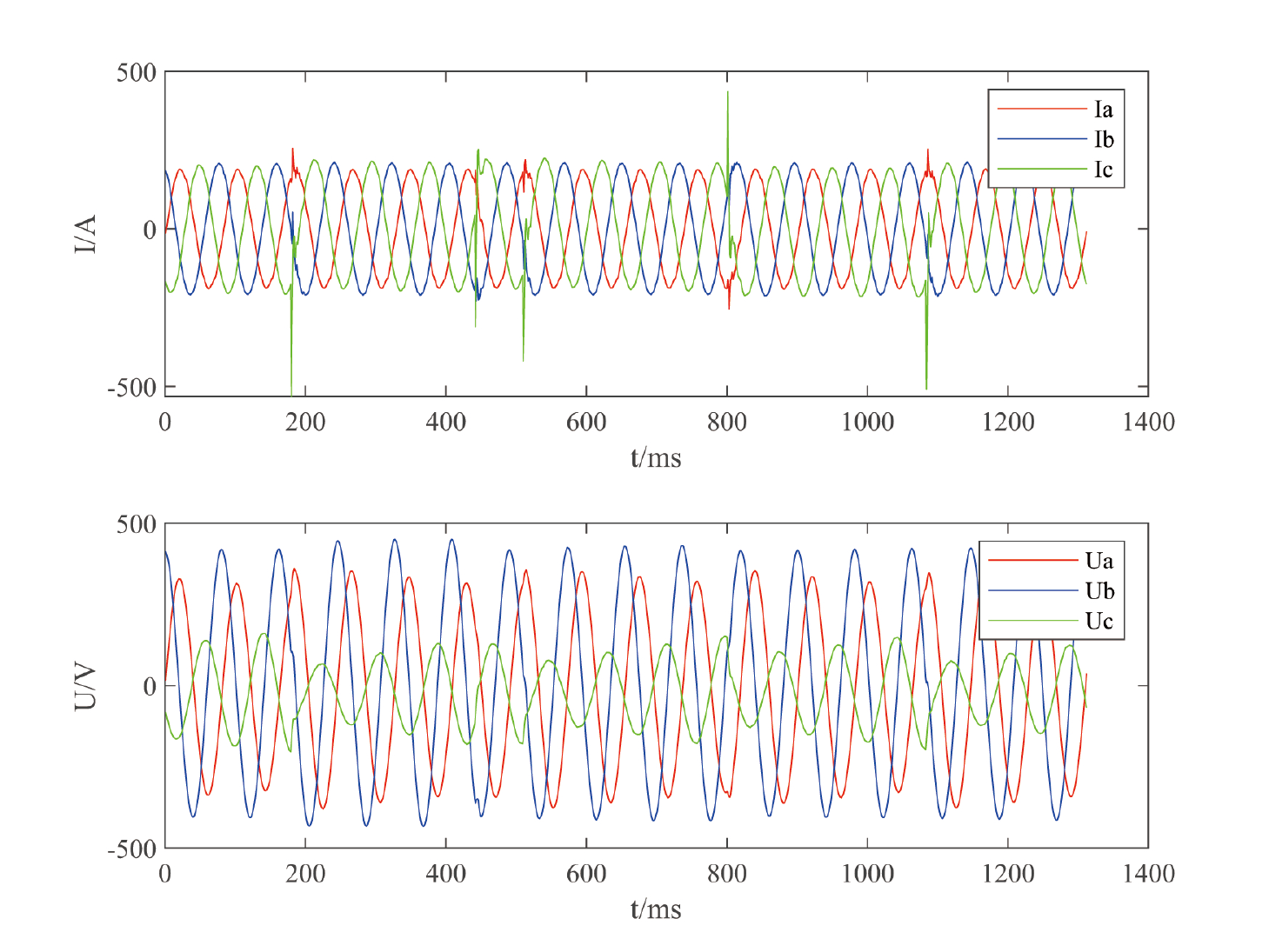}%
\label{fig_single-phase}
}
\quad
\subfloat[]{\includegraphics[width=3.4in]{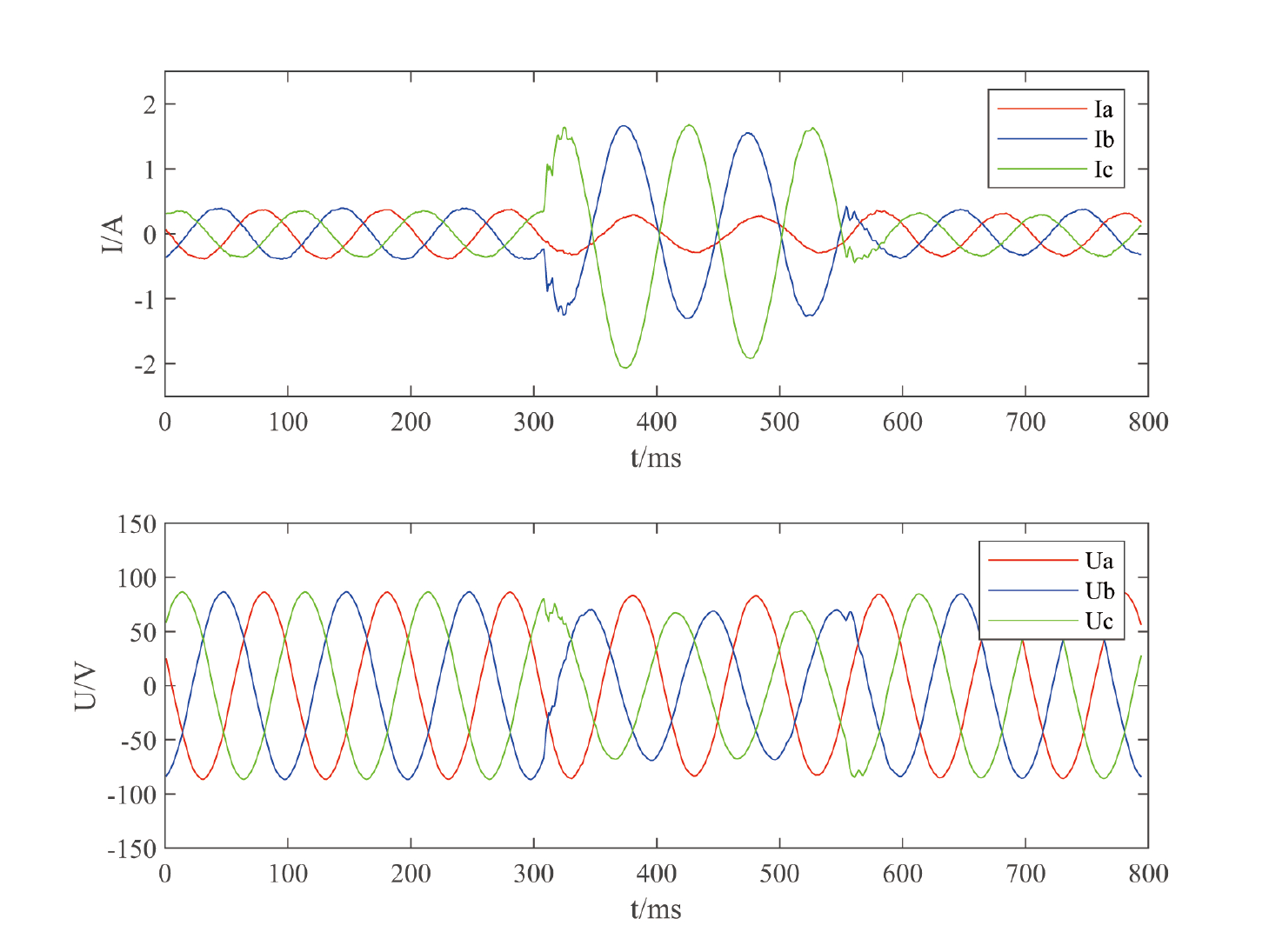}%
\label{fig_two-phase}
}
\hspace{-8mm}
\subfloat[]{\includegraphics[width=3.4in]{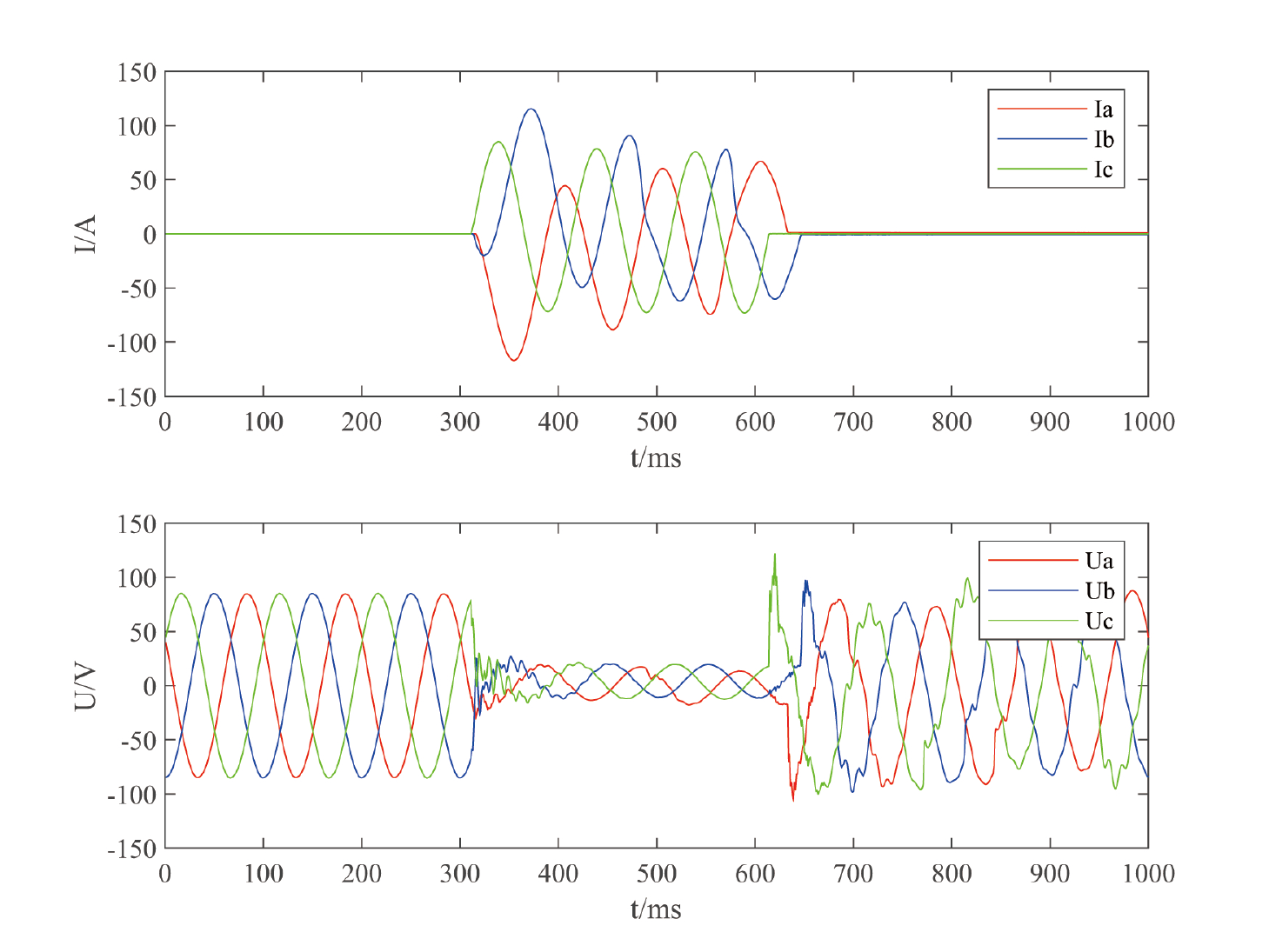}%
\label{fig_interphase}

}

\centering
\caption{Typical incipient fault signals. (a) Sub-cycle incipient fault. (b) Multi-cycle incipient fault. (c) High resistance grounding fault. (d) Single-phase grounding fault. (e) Two-phase grounding fault. (f) Interphase short circuit fault.}
\label{waveform of fault}
\end{figure*}

The faults in power distribution system, can be divided into sub-cycle faults, multi-cycle faults and permanent fault, according to their durations. Among them, sub-cycle and multi-cycle faults are called incipient faults\cite{8716299}. Sub-cycle incipient faults are characterized by abnormal fault phase voltage and recovery in one cycle, while multi-cycle incipient faults mainly include interphase short circuit faults, permanent faults include grounding with high resistance, single-phase grounding faults and main transform faults. Several typical fault waveforms are shown in the Fig. \ref{waveform of fault}.

\begin{figure*}[!t]
\centering
\subfloat[]{
\includegraphics[width=2.9in]{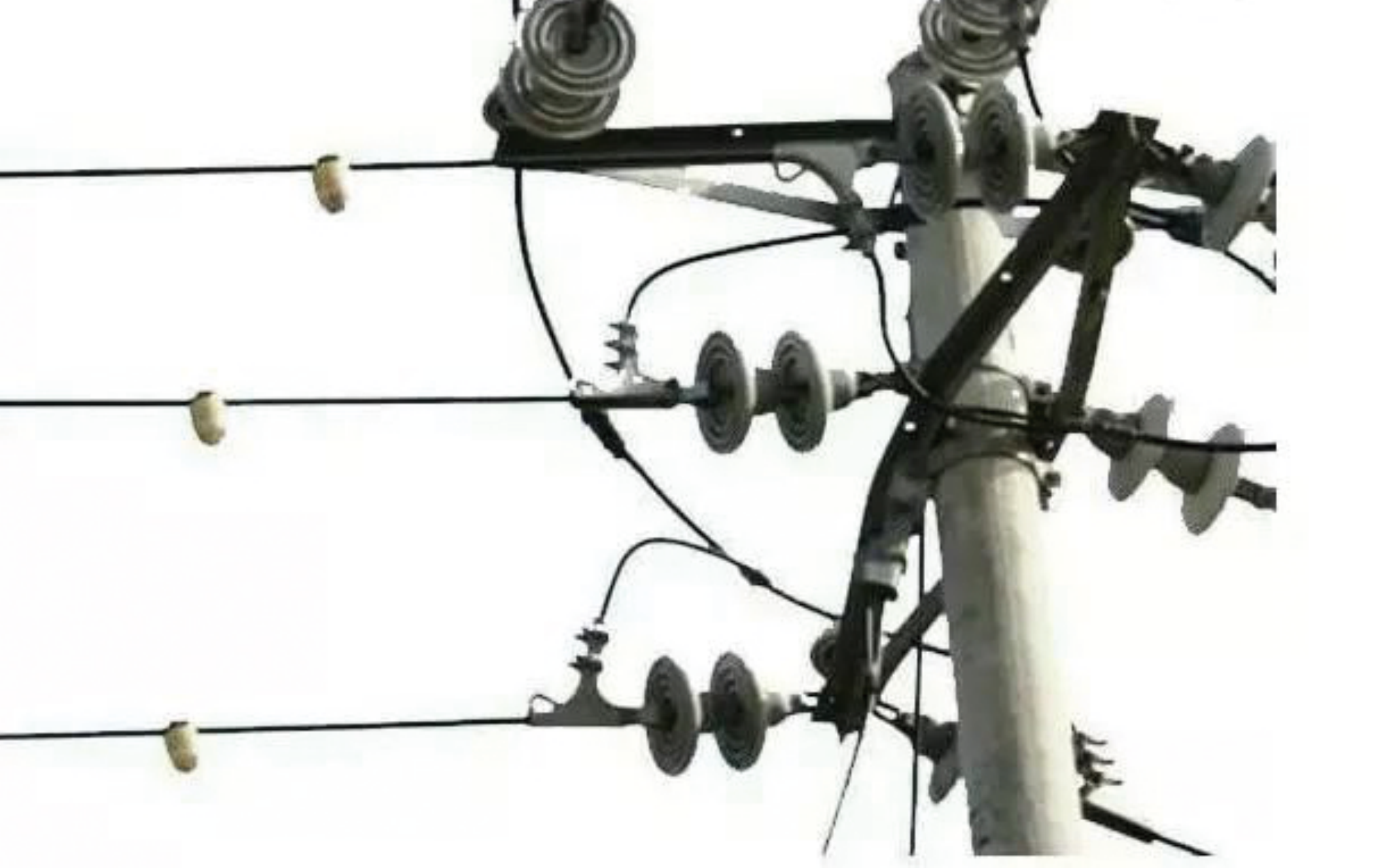}
}
\hspace{10mm}
\subfloat[]{
\includegraphics[width=1.9in]{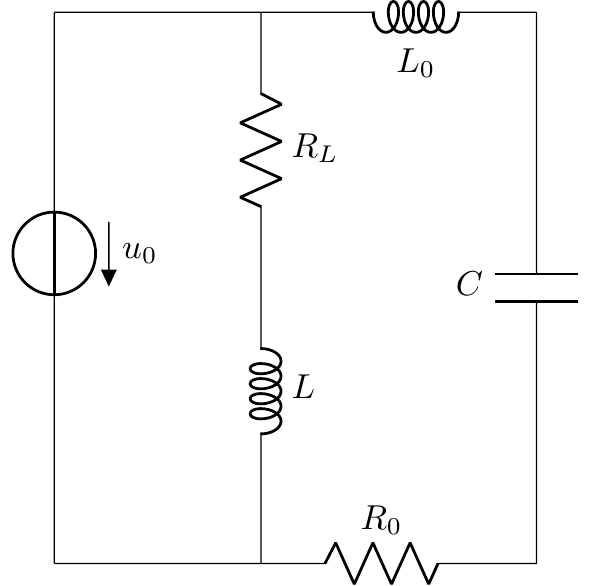}
\label{simplified_circuit}
}
\caption{Overhead line in power distribution network and its simplified circuit model. (a) Overhead line. (b) Simplified circuit model.}
\label{fig_simplified_circuit_model}
\end{figure*}

Taking the single-phase grounding fault occurring in overhead lines of power distribution network as an example, the simplified circuit model contains two inductors and one capacitor, as illustrated in Fig. \ref{fig_simplified_circuit_model}. Due to the presence of inductors, the current in the line can not be changed suddenly, which will cause a short circuit transient process, and there are a large number of integer and non-integer harmonics in the voltage and current signal. As the characteristic frequency components in the transient process are not fixed, the current signal flowing through the capacitor contains fault information, and is non-stationary. 

\subsection{Transient capacitive current} 

Based on Fig. \ref{simplified_circuit}, the differential equation of transient capacitance current can be expressed as:
\begin{equation} \label{Stationary signals4}
\begin{aligned}
R_{0}i_{C}+L_{0}\frac{di_{C}}{dt}+\frac{1}{C} \int_0^t i_{C} =  U_{m}\sin(\omega t+\varphi),
\end{aligned}
\end{equation}

\noindent where $U_{m}$ is the amplitude of the zero sequence voltage. The transient capacitive current $i_{C}$ is composed of transient free oscillation component $i_{C.os}$ and steady-state power frequency component $i_{C.st}$. When single-phase grounding fault occurs, $i_{C.os}+i_{C.st}=0$ and $i_{Cm}=U_{m}\omega C$. And the transient capacitive current can be calculated as:
\begin{equation} \label{Stationary signals5}
\begin{aligned}
i_{C}&=i_{C.os}+i_{C.st}  \\
&=I_{Cm}[\left(\frac{\omega_{f}}{\omega}\sin\varphi\sin\omega_{f}t-\cos\omega\cos\omega_{f}t\right)e^{-t\delta}+\\
~&\cos\omega_{f}t],
\end{aligned}
\end{equation}

\noindent where $I_{Cm}$ is the amplitude of the transient capacitor current. $\omega_{f}$ is the angular frequency of the transient free oscillation component. $\delta={1/\tau C}={R/2L_{0}}$ is the attenuation coefficient of the free oscillation component. $\varphi$ is the phase angle of the phase voltage when the fault occurs.

If $R_{0}$ is less than $2\sqrt{L_{0}/C}$, the transient process of the loop current has periodic oscillation and attenuation characteristics. Otherwise, the loop current has aperiodic oscillation attenuation characteristics, and gradually tends to be in a stable state.

\subsection{Transient inductive current}
The inductive current of arc suppression coil is composed of transient DC component and steady-state AC component, which is expressed as:
\begin{equation} \label{Stationary signals6}
\begin{aligned}
i_{L}=I_{Lm}[\cos\varphi^{-t/\tau_{L}}-\cos(\omega t+\varphi)],
\end{aligned}
\end{equation}
where $\tau_{L}$ is the time constant of the inductance circuit. $I_{Lm}=\frac{U_{m}}{\omega_{L}}$, $\phi$ is the phase angle of phase voltage at fault.

The fault signal contains a large number of non-fundamental transient signals, which consisting of high frequency components, non-periodic components and a large number of fault or disturbance information. The transient component in fault signal is a non-stationary random process, which changes with time, the location of the fault point, the transition resistance of the fault point and the different operating conditions of the system. Through the above analysis, when a single-phase grounding fault occurs in the power grid, it can be seen from the analysis of the transient process that the fault signal is non-stationary at that time.

\section{Time-Frequency Memory Cell Based on Adaptive Wavelet}
\label{Wavelet_based_LSTM_cell}
To extract the dynamic characteristics of fault parameters in power distribution networks, we introduce wavelet transform that can accurately analyze non-stationary signals into the LSTM cell, and change the forget gate of LSTM into the joint forget gate, which decomposes fault information in both time and frequency domain. 
Besides we establish an adaptive learning mechanism for scale parameters and position parameters in wavelet transform, and propose AD-TFM cell that can accurately model the non-stationary incipient fault signal.

\subsection{Basic Idea of AD-TFM}
The traditional method, which combines the wavelet transform and the neural network for fault detection, usually uses the wavelet transform to extract the fault features, which are then fed into neural network for classification\cite{9177071}. In this method, the wavelet transform is separated from the neural network, and the error generated during feature extraction has a greater impact on the accuracy of fault classification in the later stage. To solve the above-mentioned problem, we propose AD-TFM by embedding the wavelet transform into traditional LSTM cell, changing the originally fixed scale parameter and translation parameter to dynamic parameter that changes with the input fault information.

\subsection{Structure of AD-TFM Cell}
\begin{figure*}[htb!]
    \centering
    \includegraphics[width=6.1in]{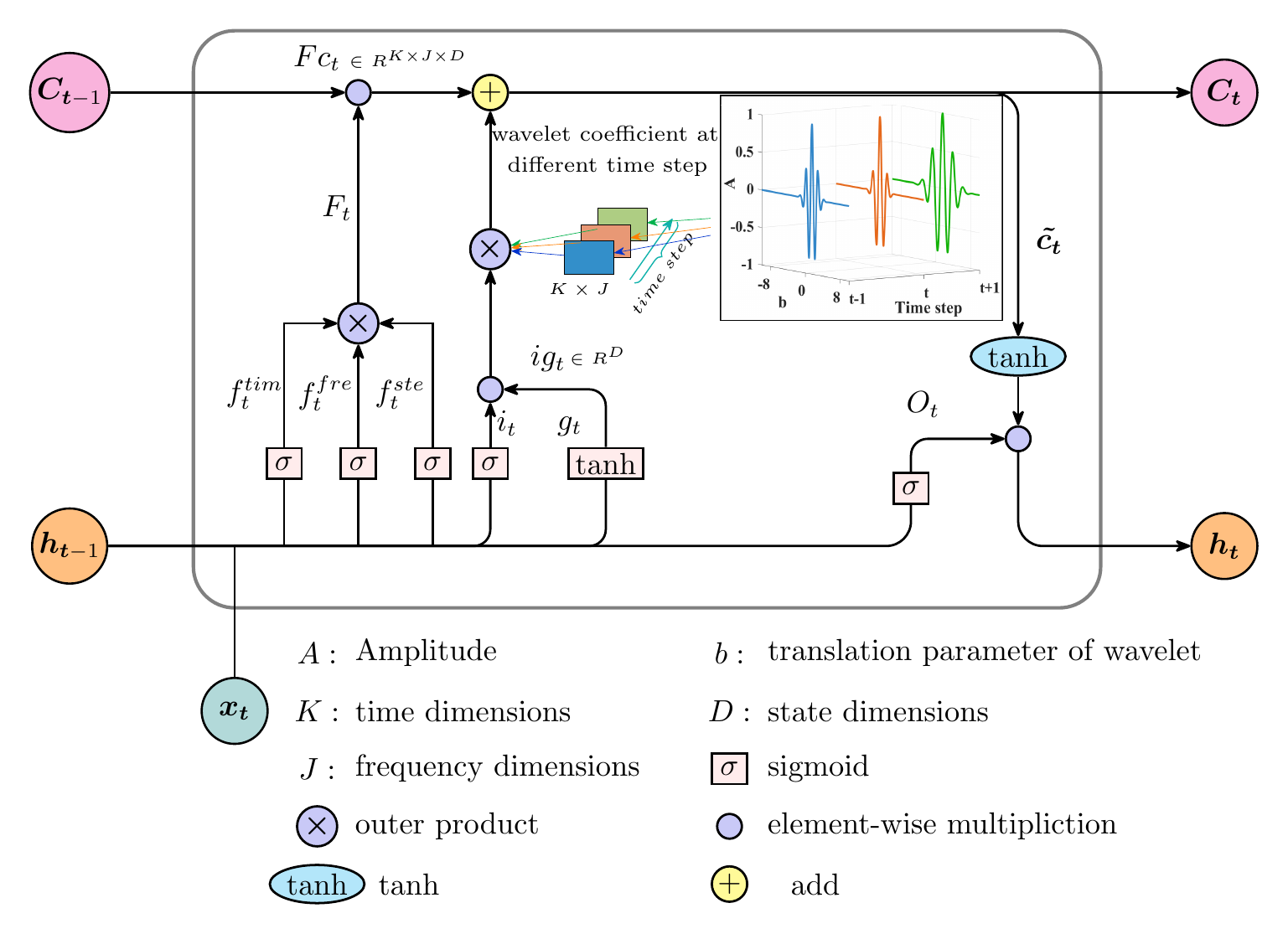}
    \caption{Structure of AD-TFM cell.}
    \label{fig_tfmcell}
\end{figure*}

The structure of AD-TFM cell is shown in Fig. \ref{fig_tfmcell}, which consists of joint forget gate, input gate, output gate and cell state updating.
It dynamically models the input, i.e., three-phase current and voltage time series $\left\{x_{t}\mid t = 1,2,...,T\right\}$ by continuous time steps. In each time step of AD-TFM, the hidden state of the previous time step and the input information of the current time step are decided by the joint forget gate, and the input gate selects the information to be updated. In the cell state updating part, the input information after adaptive wavelet transform and the information retained by the joint forget gate are added to update the cell state, and then the updated cell state is input to the output gate to obtain the hidden state at the current time.

In this process, the non-stationary analysis is achieved by converting the input three-phase voltage and current data into time-frequency features using efficient time modeling (via LSTM) and  non-stationary signal processing (i.e., wavelet transform).

\subsection{The Joint Forget Gate}
\label{The Joint Forget Gate}
The joint forget gate contains three parts: the state forget gate $f_t^{ste}$, the time forget gate $f_t^{tim}$ and the frequency forget gate $f_t^{fre}$, which decompose the input and the hidden state of the previous time step into the $K$ dimension in the time domain, $J$ dimension in the frequency domain and $D$ dimension in state domain, respectively.

\begin{equation} \label{state forget gate}
\begin{aligned}
f_t^{ste} = sigmoid\left(W_{ste}x_{t} +U_{ste}h_{t-1} + b_{ste} \right)\in \mathbb{R}^{D},
\end{aligned}
\end{equation}

\begin{equation} \label{time forget gate}
\begin{aligned}
f_t^{tim} = sigmoid\left(W_{tim}x_{t} +U_{tim}h_{t-1} + b_{tim} \right)\in \mathbb{R}^{K},
\end{aligned}
\end{equation}

\begin{equation} \label{frequence forget gate}
\begin{aligned}
f_t^{fre} = sigmoid\left(W_{fre}x_{t} +U_{fre}h_{t-1} + b_{fre} \right)\in \mathbb{R}^{J},
\end{aligned}
\end{equation}

\noindent where $W_{*}$ and $U_{*}$ are weight matrices. $b_{*}$ is a bias vector and $h_{t-1}$ is the output hidden state at the ($t-1$)th time step. Among them ${*}$ refers to ${ste}$, ${tim}$, ${fre}$.

The output of three forget gates are used to obtain $F^{t}$, by jointly regulating the state, time and frequency information.
\begin{equation} \label{joint forget gate}
\begin{aligned}
F_{t} = f_t^{ste} \otimes f_t^{tim} \otimes f_t^{fre}
\in\mathbb{R}^{D\times J\times K},
\end{aligned}
\end{equation}

\begin{equation} \label{state1}
\begin{aligned}
FC_{t}=F_{t}\circ C_{t-1}
\in\mathbb{R}^{D\times J\times K},
\end{aligned}
\end{equation}

\noindent where $\otimes$ is the outer product operation and $\circ$ is the element-wise multiplication operation. 

The joint forget gate determines the amount of information retained from the previous time step to the current step. It can be considered as a combination gate, which controls the information of different frequencies, times and states flowing into the memory cell.

\subsection{Input Gate}
\label{Input Gate}

The formulations of the input gate $i_{t}$ and the input modulation $g_{t}$ are similar as these of LSTM:

\begin{equation} \label{input gate}
\begin{aligned}
i_{t} = sigmoid\left(W_{i}x_{t}+U_{i}h_{t-1}+b_{i}\right),
\end{aligned}
\end{equation}

\begin{equation} \label{input modulation}
\begin{aligned}
g_{t} = tanh\left(W_{g}x_{t}+U_{g}h_{t-1}+b_{g}\right),
\end{aligned}
\end{equation}

\begin{equation} \label{input }
\begin{aligned}
ig_{t} = i_{t}\circ g_{t},
\end{aligned}
\end{equation}
\noindent where the $ig_{t}$ is defined to generate a compatible result for the input gate.

The input gate decides how much new information should be allowed to enter the current memory cell to update AD-TFM.

\subsection{Cell state updating based on adaptive wavelet transform}
\label{Updating cell state}
The state updating procedure of AD-TFM is similar to LSTM. By integrating the adaptive wavelet transform, the output of the input gate needs to be multiplied by the coefficients of the adaptive wavelet transform when the AD-TFM cell is updated.
The output of the input gate is decomposed by the wavelet transform into $K$ and $J$ dimensions in the time domain and frequency domain respectively.

Taking the Morlet wavelet transform used in this paper as an example, the implementation function of the adaptive learning of the scale parameters $a$ and translation parameters $b$ are:
\begin{equation} \label{ Morlet wavelet2}
\begin{aligned}
a =tanh\left(W_{a}ig_{t}+b_{a}\right),
\end{aligned}
\end{equation}
\begin{equation} \label{ Morlet wavelet3}
\begin{aligned}
b =tanh\left(W_{b}ig_{t}+b_{b}\right).
\end{aligned}
\end{equation}

The output of the input gate, which is decomposed by the wavelet transform, is expressed as follows.

\begin{equation} \label{Morlet wavelet}
\begin{aligned}
\psi_{k,j} &=exp\left(i\cdot \frac{\omega_{0}}{a}\cdot\left(\frac{t+b}{2^{j}}-k\right)\right)\cdot\\
~& exp\left(\left(-\frac{1}{a}\right)\cdot\left(\frac{t+b}{2^{j}}-k\right)^{2}\right).
\end{aligned}
\end{equation}

Then the cell state after decomposition can be obtained as:

\begin{equation} \label{decomposition cell state}
\begin{aligned}
C_{t}=FC_{t}+ig_{t}\otimes \psi_{k,j}\in\mathbb{R}^{D\times J\times K},
\end{aligned}
\end{equation}
\noindent $C_{t-1}$ is the cell state value at the previous time step. $FC_{t}\in\mathbb{R}^{D\times J\times K}$ and $i_{t}\in\mathbb{R}^{D}$ are forget and input gates, respectively, controlling the past and current information on states, time and frequencies that are allowed to update the AD-TFM at the $t$th time step.

As a complex number can be uniquely represented by its amplitude and phase, we decompose the update matrix $C_{t}$ of AD-TFM into two parts, amplitude and phase, which are expressed as:
\begin{equation} \label{amplitude}
\begin{aligned}
A_{t}=\mid C_{t}\mid=\sqrt{\left(Re C_{t}\right)^{2}+\left(Im C_{t}\right)^{2}} \in\mathbb{R}^{D\times J\times K},
\end{aligned}
\end{equation}

\begin{equation} \label{phase}
\begin{aligned}
\angle C_{t} = arctan\left(\frac{Re C_{t}}{Im C_{t}}\right)\in\left[-\frac{\pi}{2},\frac{\pi}{2}\right].
\end{aligned}
\end{equation}

\noindent where $Re$ and $Im$ are the functions of taking the real part and taking the imaginary part, respectively. $arctan\left(\cdot\right)$ is an element-wise inverse tangent function.

The amplitude will be fed into the memory cell for the next time step. However,
this phase will be ignored because it has no impact on performance except for higher computation and memory overhead.

At each time step, we calculate the component $A_t^{k,j}$ of the amplitude $A_{t}$ in the $k$th dimensional time domain and the $j$th dimensional frequency domain. $A_t^{k,j}$ will be sent to next time step state cell unit of AD-TFM, and the forget gate and input gate determine the information that needs to be updated. After the update, $A_t^{k,j}$ is combined into $\widetilde{c_{t}}$, and enters the output gate, which is expressed as:

\begin{equation} \label{Combination_A}
\begin{aligned}
\widetilde{c_{t}}=\sum\nolimits_{k=1}^K\sum\nolimits_{j=1}^J\left(W_c^{k,j}A_t^{k,j}+b_c^{k,j}\right),
\end{aligned}
\end{equation}

\subsection{Output gate}
\label{Output gate}

The output gate determines the information that will be fed into next time step. The input of the output gate can be expressed as:

\begin{equation} \label{output_gate_o}
\begin{aligned}
o_{t} = sigmoid\left(W_{o}x_{t} +U_{o}h_{t-1} + b_{o} \right),
\end{aligned}
\end{equation}

And the output hidden state $h_t$ is computed as:
\begin{equation} \label{output_gate_h}
\begin{aligned}
h_{t}=o_{t}\circ tanh\left(\widetilde{c_{t}}\right).
\end{aligned}
\end{equation}

\section{AD-TFM Based RNN with Attention for Incipient Fault Detection}
\label{Fault_Detection_Model}
\begin{figure*}[htb!]
    \centering
    \includegraphics[width=5.5in]{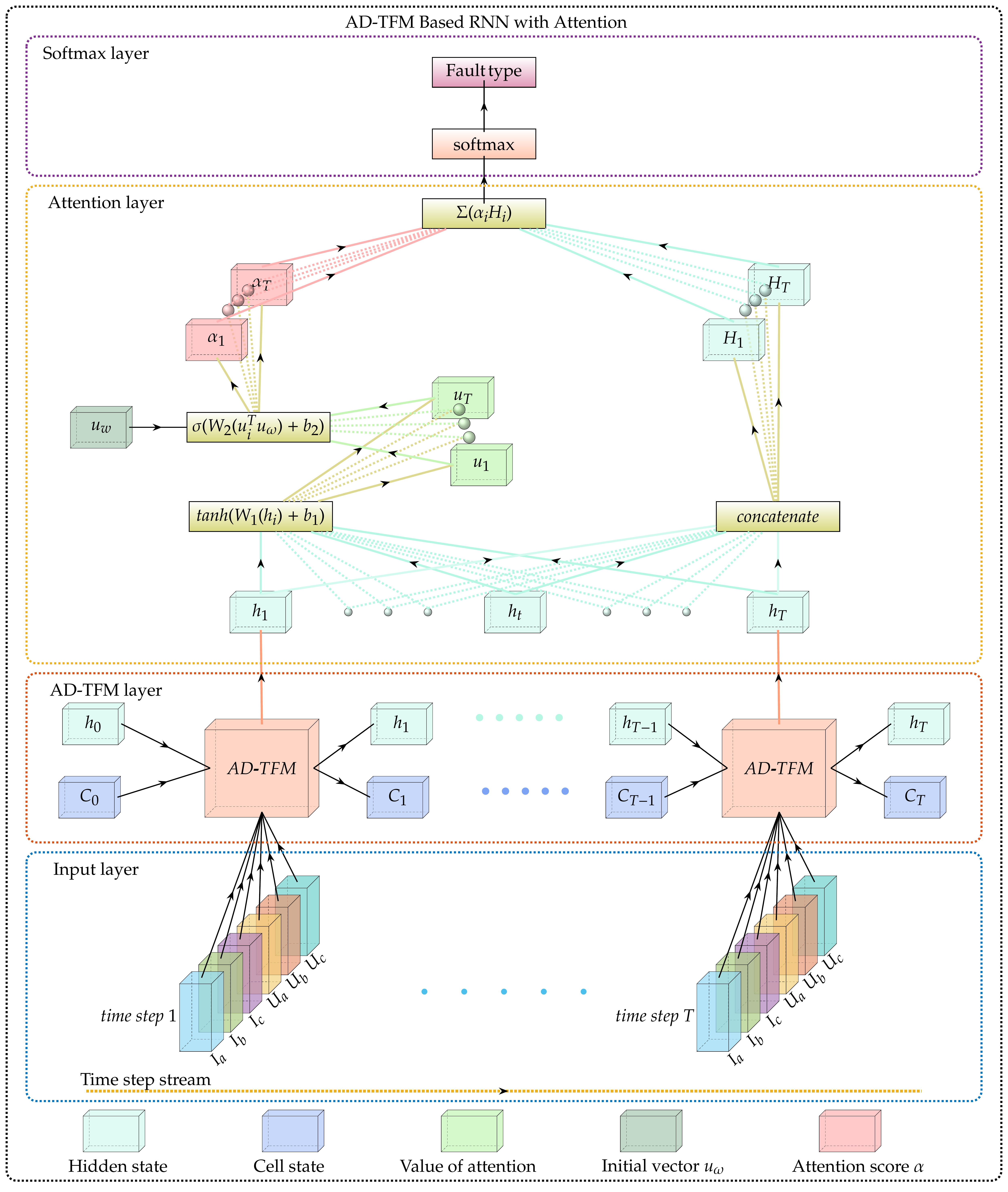}
    \caption{AD-TFM-AT model for incipient fault detection.}
    \label{fig_muti_tfm}
\end{figure*}

In this section, we construct a RNN model for incipient fault detection based on the proposed AD-TFM cell. 
To focus the neural network on the global hidden information, we strengthen the stacked AD-TFM network by adding an attention layer.
The hierarchical structure of AD-TFM-AT model is shown in Fig. \ref{fig_muti_tfm}.

The fault signal consisting of three phase voltage and current input to AD-TFM cell will be encoded into a fixed length of hidden information. At each time step, the hidden state output contains a different amount of fault information. Direct use of the hidden state of last time step of AD-TFM will lead to insufficient attention to global hidden information. Therefore, the amount of fault information contained in the hidden information of each time step of AD-TFM cell is quantified in the form of similarity through the attention mechanism, and the final output is calculated by weighting the matching degree of hidden state of each time step as a weight. Therefore, we use the attention mechanism to extract the important information from the hidden states of  all time steps and give it larger weights to obtain more accurate fault feature vectors and improve the fault detection accuracy. Its specific implementation process is as follows:

Let ${h_{i}}$ represent the hidden layer vector containing the time series produced by AD-TFM. We convert ${h_{i}}$ to ${u_{i}}$ through a fully connected layer illustrated as: 
\begin{equation} \label{attention1}
\begin{aligned}
u_{i} = tanh\left(Wh_{i} + b_{o} \right).
\end{aligned}
\end{equation}

Then we calculate the similarity between $u_{i}$ and the context vector $u_{w}$, and convert it to a probability distribution $\alpha_{i}$ through softmax function.
\begin{equation} \label{attention_alpha_i}
\begin{aligned}
\alpha_{i} = \frac{exp(u_{i}^{T}u_{w})}{\Sigma_{i}exp(u_{i}^{T}u_{w})}.
\end{aligned}
\end{equation}

The context information $u_w$ can be regarded as the contribution of one time step data to the overall data, and the contribution of each $u_i$ to $u_w$ can be obtained by calculating the similarity between $u_i$ and $u_w$, where $u_w$ is randomly initialized and obtained through training.

As $\alpha_i$ represents the importance of the fault hidden state at each time step to the overall fault hidden state, we use $\alpha_i$ as the weighted summation of the global $h_i$ to obtain the tensors and express the fault type.
\begin{equation} \label{attention_s}
\begin{aligned}
s= \Sigma\alpha_{i}h_i.
\end{aligned}
\end{equation}

\section{Data augmentation}
\label{Data_augmentation}

\subsection{Overview}
The incipient faults of the power distribution systems are manifested as waveform distortion of the three-phase voltage and current sinusoidal signals at the moment of fault occurrence. We intercept the three-phase voltage and current data before and after the moment of fault occurrence as fault data. In Section \ref{Analysis of initial fault signals}, we have introduced the types of incipient fault signals in the power distribution system and showed the waveform of the faults.

Orthogonally, training a neural network usually requires a large amount of data. However, low incidence of incipient fault makes it a typical small sample learning problem \cite{9094224}. According to the characteristics of voltage and current sinusoidal signal, we use two methods for data augmentation, i.e., phase switching and temporal sliding, to obtain a larger training dataset while keeping the characteristics of fault data unchanged.

\subsection{Phase Switching}
\label{Switch Phase}

The single-phase grounding fault is one major incipient fault in power distribution systems, where the faults happens in one phase of the three-phase voltage and current data. 
The first data augmentation method we use is phase switching, which swaps the phase of the voltage and current data of the fault signal with one of the rest phases. This changes the phase of the fault but keeps the fault type unchanged, i.e., achieve multiple data from one fault data.

Taking a single-phase grounding fault as an example, we assume that the fault occurs in phase A, i.e., phase A voltage and current data contains fault information, and phase B and C voltage and current data are normal. Then, we swap the voltage and current data of the fault occurring in phase A with the normal data of phase B. In this way, the fault occurring in phase A becomes the fault occurring in phase B, and new data containing fault can then be obtained.
Meanwhile, this operation does not change the characteristics of the single-phase grounding fault, e.g., the fault does not happen in two or more phases at the same time.
We also exchange the voltage and current data of the fault occurring in phase A with the normal data of phase C.

\subsection{Temporal Sliding}
\label{Time Domain Expansion}
Temporal sliding is also used to enlarge the amount of fault data, by sampling the original data containing the fault data multiple times with different starting time.
The starting time here is selected with equal time sliding intervals.
Comparing with original data (i.e., the case of only one starting time), the amount of fault data is increased while the characteristics of fault data is unchanged.

\begin{figure}[t!]
    \centering
    \includegraphics[width=3.2in]{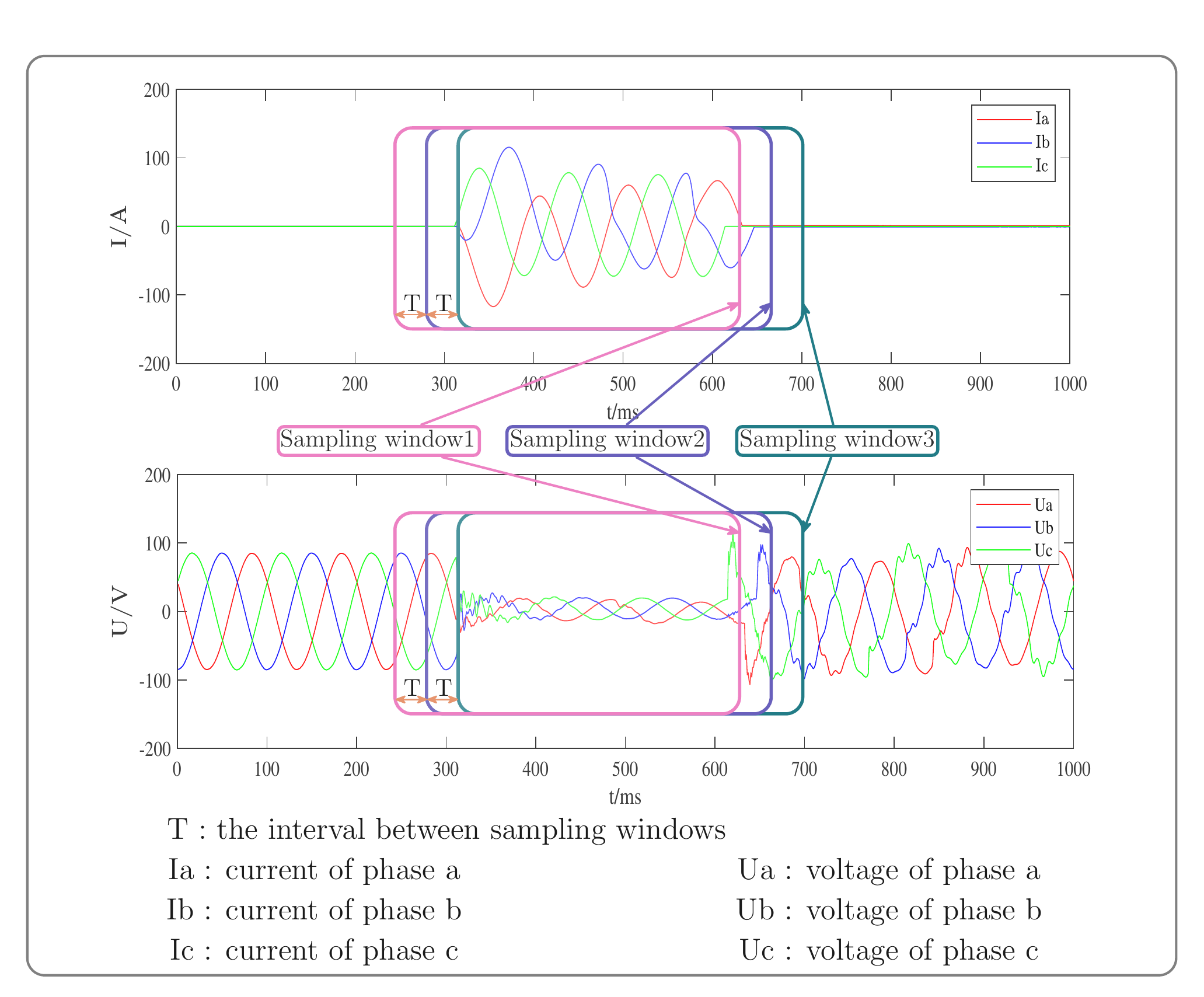}
    \caption{Illustration of temporal sliding.}
    \label{Temporal sliding}
\end{figure}

In particular, we select a window of a certain length H to intercept the fault data, and the window can pick different starting points when sampling the fault data. In order to enlarge the amount of fault data, we specify that the sampling window intercepts the data from one starting time point and then slides backward $T$ time points to intercept the data again. In this way, a fault can be intercepted multiple times and the amount of fault data is increased, thus achieving data augmentation.
Fig. \ref{Temporal sliding} illustrates the temporal sliding. 
With sliding windows at different starting times, one fault data will be sampled multiple times within different sliding windows. As a result, the amount of fault data increases, but the type of fault is not changed.

\section{Experiments}
\label{sec_Experiments}
To verify the performance of our proposed AD-TFM-AT neural network model, extensive experiments are conducted on two datasets.
We use several evaluation metrics to assess the performance with and without data enhancement.
We also perform ablation experiments to show the performance of adaptive wavelet transform and attention mechanism.

\subsection{Experimental Setup}
\textbf{Dataset and Analysis:} To train and test the proposed model, we use two datasets, a small Incipient Fault dataset in Power Distribution (IFPD) system from \cite{9094224}, and a relatively large dataset logged by State Grid Corporation of China in AnHui Province (SGAH). In IFPD dataset, there are the Sub-cycle Incipient Fault (SIF), Multi-cycle Incipient Fault (MIF), Single phase Grounding Fault (SGF) and High Resistance Grounding Fault (HRGF), each containing three-phase voltage and three-phase current data, and each cycle has 82 sampling points. The waveforms are shown in Fig. \ref{fig_subcycle}, Fig. \ref{fig_multicycle}, Fig. \ref{fig_single-phase} and Fig. \ref{fig_high}, respectively. In the SGAH dataset, there are Inter Phase Short-circuit Fault (IPSF), Two-phase Ground Fault (TGF), Single-phase Grounding Fault (SGF), and Main Transformer Fault (MTF), each also containing three-phase voltage and three-phase current data, with 100 sampling points per cycle. Both datasets contain groundtruth consisting of three-phase voltage and current data with fault labels.
The waveform of these faults are shown in Fig. \ref{fig_two-phase}, Fig. \ref{fig_interphase} and Fig. \ref{fig_single-phase}. We also make SGAH dataset available to the public at GitHub.

\textbf{Evaluation Metrics:} To verify the performance of our proposed method, the following five metrics are adopted: accuracy, precision, recall, F1-score and Receiver Operating Characteristic (ROC) curve. To calculate the accuracy, the fault detection results are compared with ground truth. Accuracy is the ratio of the number of correct predictions to the total number of samples. Precision is the ratio of the number of samples correctly classified as faults to the number of samples in the population that are classified as such.
The higher the precision is, the better performance of the model will gain. Recall rate refers to the fact that the number of samples that are correctly classified for a certain type of fault accounts for the actual samples. The higher the recall rate is, the less the number of faults that are incorrectly classified into other types of faults will be. To balance the accuracy and recall of our model, we also calculate the F1-score. The performance of the proposed model are also evaluated by the size of Area Under ROC Curve (AUC). The larger the AUC is, the better the performance of the model has.

\textbf{Data Augmentation Evaluation:} To verify the validity of the proposed data augmentation method, both dataset are divided into original dataset and augmented dataset. Then we train the proposed AD-TFM-AT network, and test the performance.

\textbf{Ablation experiments:}  we conduct ablation experiments to show the performance of TFM, AD-TFM and TFM-AT. TFM model is based on the LSTM by changing the forget gate to a joint forget gate and adding a wavelet transform with fixed scale parameters and translation parameters. AD-TFM is based on TFM with wavelet transform of learnable parameters without the attention mechanism. TFM-AT is based on TFM with the attention mechanism.

{\textbf{Comparison schemes:}
To validate the classification performance of the proposed AD-TFM-AT, we use the following five comparision schemes. We train these models using both IFPD and SGAH datasets with augmentation, and compare the evaluation metrics on the ground truth.
\begin{enumerate}
\item Support Vector Machines (SVM): The three-phase voltage and current data are input into a set of Gaussian kernel functions based SVMs, where each SVM detects one kind of faults. And the classification results of all SVMs are combined to achieve fault classification.
\item LSTM: The pre-processed three-phase voltage and current data are fed into a three-layer LSTM for learning. Then feature classification of the LSTM output is implemented by a fully connected layer.
\item Minirocket\cite{10.1145/3447548.3467231}: Multiple features of three-phase voltage three-phase current data are extracted using multiple convolution kernels which are represented by two deterministic values $\{-1,2\}$. The multiple features are then used to train a linear classifier for fault detection.
\item SlI-CNN\cite{9702755}: The  three-phase voltage and current data are converted to synchronous Lissajous images as the input to a CNN. The CNN contains three convolutional layers and one fully-connected layer. Each convolutional layer consists of batch normalization, max-pooling, and dropout. The last convolutional layer connects a fully-connected layer for classification.
\item HLCL\cite{9094224}: The  three-phase voltage and current waveform are decomposed into approximate shapes and residuals by Meyer wavelet, and then further decomposed into primitive and temporal relationships by Fast Fourier Transform (FFT). Finally, fault classification is realized by variable probability statistics and Bayesian hierarchical model. 
\end{enumerate}
}

\textbf{Implementation details:} The proposed model is implemented in Python3.7, and the experimental code is available at GitHub\footnote{ \url{https://github.com/smartlab-hfut/AD-TFM-AT-Model}}. The training parameter settings are shown in Table~\ref{table_implement_params}. All the experiments are performed on four Nvidia Tesla V100 GPUs. We use a cycle of three-phase voltage and current data as a data packet.

\begin{table}[htb!]
\captionsetup{justification=centering}
    \renewcommand\arraystretch{1}
 \centering
 \caption{Experimental parameter settings.}
 \label{table_video_params}
 \centering
    \setlength{\tabcolsep}{0.5mm}{
 \begin{tabular}{c c c c c c}
    \toprule
    \toprule
    Network model& LSTM   & TFM &AD-TFM & TFM-AT &  AD-TFM-AT \\ 
    \midrule
    Initial learning rate & 0.001  &0.001 &0.001& 0.001 &0.001   \\
    Batch size& 256  & 256 & 256 & 256 & 256 \\
    Optimizer &Adam  & Adam& Adam& Adam& Adam \\ 
     $D$      &32    &32   &32   &32   &32 \\
     $K$      &     &4    &4    &4    &4 \\
     $J$      &      &4    &4    &4    &4\\
    $\omega_{0}$  &     &16   &16   &16   &16\\
    \bottomrule
    \end{tabular}}
    \label{table_implement_params}
\end{table}

\subsection{ Experimental Results}
\textbf{Data augmentation results:} First we perform data augmentation on IFPD and SGAH datasets, and the results are shown in Table~\ref{data enhance}.

\begin{table}[htb!]
\captionsetup{justification=centering}
\renewcommand\arraystretch{1}
 \centering
 \caption{Augmentation of IFPD Dataset and SGAH Dataset.}
 \label{data enhance}
 \centering
    \setlength{\tabcolsep}{1mm}{
\begin{tabular}{c c c c}
\toprule
\toprule
Dataset&Fault type&Original data size&Augmented data size\\
\midrule
\multirow{5}{*}{IFPD}&SIF &137 & 3276 \\
\multirow{5}{*}{}& MIF & 164 & 2340\\ 
\multirow{5}{*}{}& SGF & 678 & 7767\\ 
\multirow{5}{*}{}& HRGF & 28 & 486\\ 
\multirow{5}{*}{}& Normal & 300 & 2700\\ 
\midrule
\multirow{5}{*}{SGAH}&IPSF &342 & 2052 \\
\multirow{5}{*}{}& TGF & 70 & 420\\ 
\multirow{5}{*}{}& SGF & 501 & 3006\\ 
\multirow{5}{*}{}& MTF & 497 & 2982\\ 
\multirow{5}{*}{}& Normal & 320 & 1920\\
\bottomrule
    \end{tabular}}
\end{table}

\begin{figure*}[htb!]
\centering
\subfloat[]{

\includegraphics[width=2.7in]{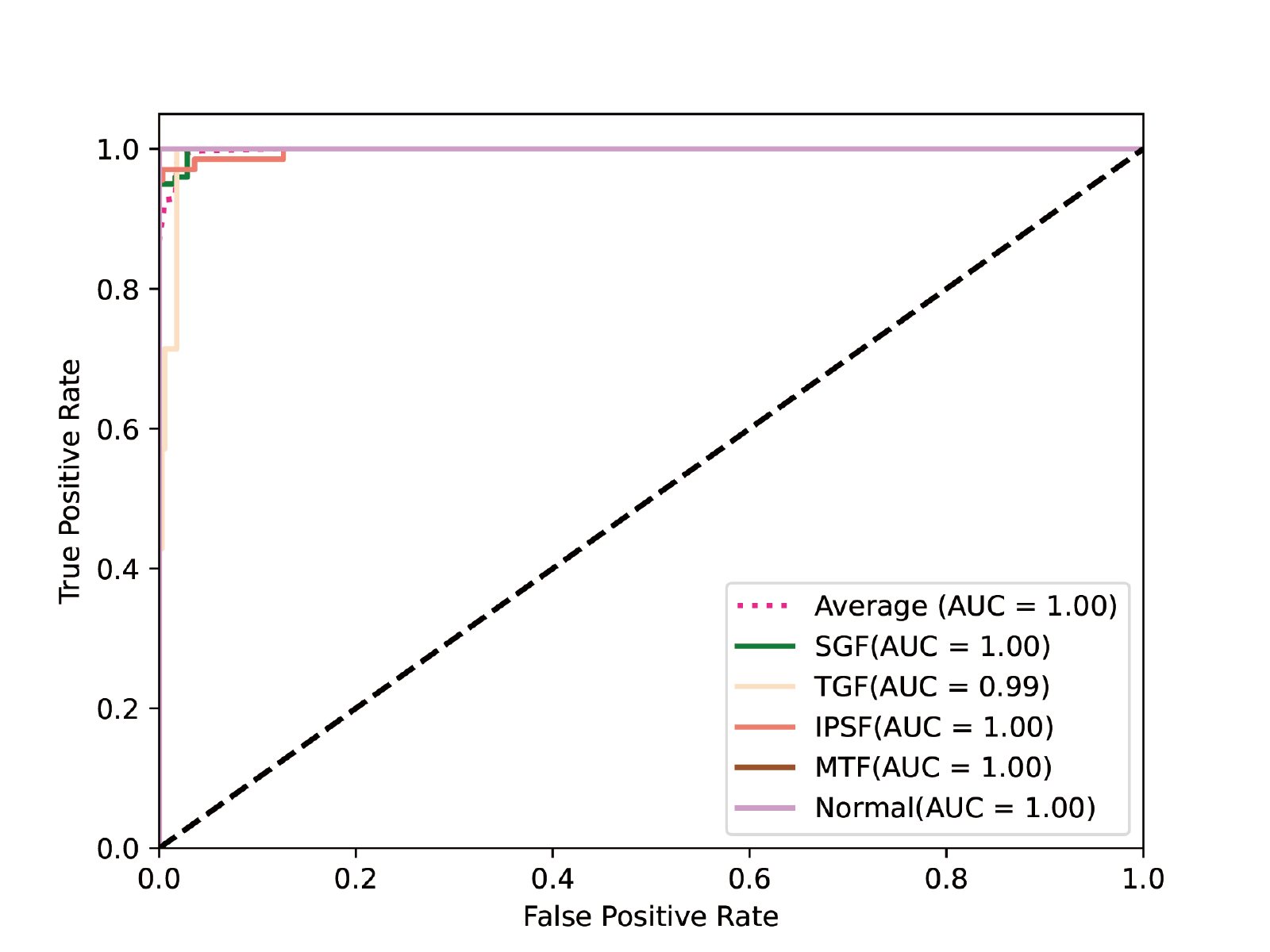}
}
\hspace{-8mm}
\subfloat[]{
\includegraphics[width=2.7in]{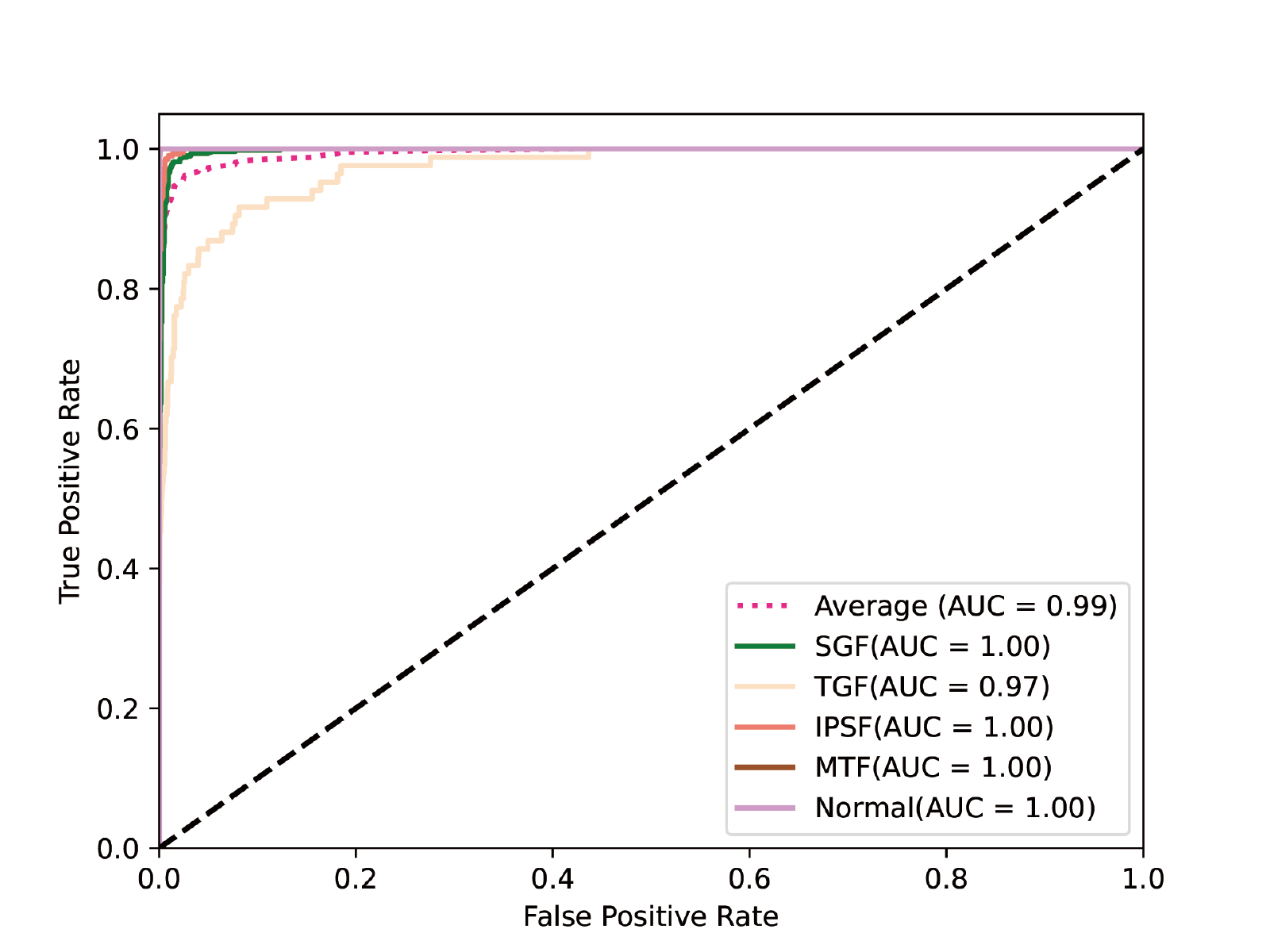}

}

\caption{ROC of AD-TFM-AT model on SGAH dataset. (a) SGAN original dataset. (b) SGAH augmented dataset.}
\label{ROC of GW ag}
\end{figure*}

\begin{figure*}[htb!]
\centering

\subfloat[ ]{

\centering
\includegraphics[width=2.7in]{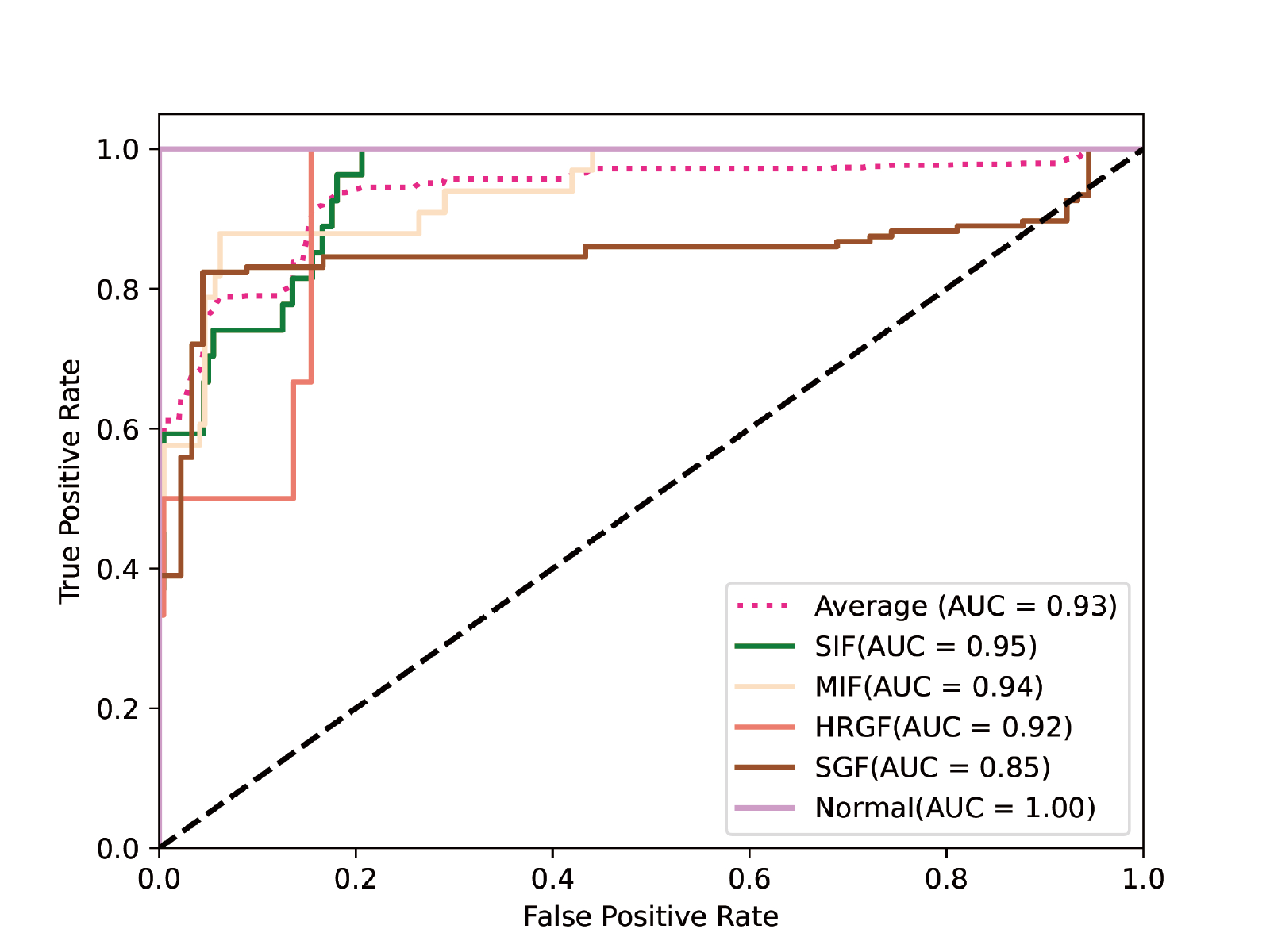}

}
\hspace{-8mm}
\subfloat[]{

\centering
\includegraphics[width=2.7in]{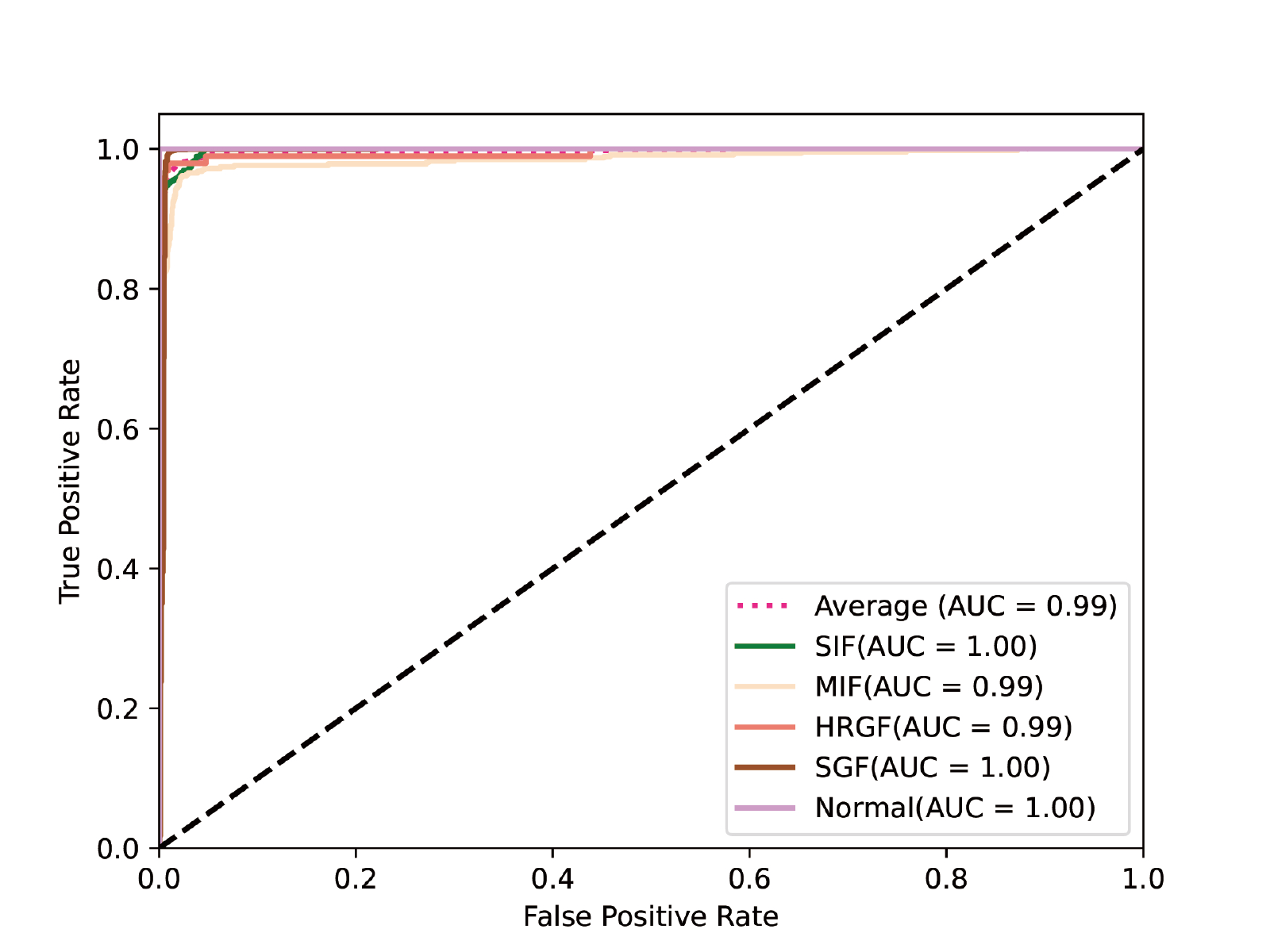}

}

\caption{ROC of AD-TFM-AT model on IFPD dataset. (a) IFPD original dataset. (b) IFPD augmented dataset.}
\label{ROC of WL ag}
\end{figure*}

Then we select the original data and augmented data on a pro-rata basis (8:2) to train and test our proposed AD-TFM-AT neural network. The performance is shown in the Table~\ref{metrics of data enhance}. And the model classification performance is represented by ROC curves in Fig. \ref{ROC of GW ag} and Fig. \ref{ROC of WL ag}.

\begin{table}[htb!]
\captionsetup{justification=centering}
\renewcommand\arraystretch{1}
 \centering
 \caption{Performance evaluation of AD-TFM-AT with augmented data.}
 \label{metrics of GW ag}
 \centering
    \setlength{\tabcolsep}{1mm}{
 \begin{tabular}{c cccc}
\toprule
\toprule
\multirow{2}{*}{Metrics}&
\multicolumn{2}{c}{SGAH Data}&\multicolumn{2}{c}{IFPD Data}\cr\cline{2-5}&Original &Augmented &Original &Augmented \\
\midrule
Accuracy & 0.97 & 0.99 & 0.82 & 0.97 \\ 
Precision& 0.96 & 0.97 & 0.83 & 0.97 \\ 
Recall   & 0.98 & 0.98 & 0.74 & 0.96 \\
F1-score & 0.97 & 0.98 & 0.75 & 0.96 \\
\bottomrule
    \end{tabular}}
    \label{metrics of data enhance}
\end{table}

From Table~\ref{metrics of data enhance}, we can see that the model trained on the augmented dataset generally performs better in terms of the accuracy, precision, and F1-score than the model trained on the original dataset. In particular, the AD-TFM-AT model trained on the augmented IFPD data better than trained on the original dataset, of which the accuracy is increased from 0.82 to 0.97, and the F1-score is increased from 0.75 to 0.96. The above experimental results show that our proposed data augmentation method provides an effective support for model training, and solves the problem that the amount of incipient fault data of the power distribution network is insufficient.

The ROC of the data augmentation experiments are shown in Fig. \ref{ROC of GW ag} and Fig. \ref{ROC of WL ag}.
We can see that the proposed AD-TFM-AT model has better classification ability on both the original and augmented SGAH datasets. In particular, the average AUC of the AD-TFM-AT model improves from 0.93 to 0.99 in the IFPD dataset, and the AUC of identifying SGF improves from 0.85 to 1. This indicates that the proposed data augmentation method allows the model to more sufficiently learn each fault characteristic and improves the model's resolution ability to various faults.

\begin{figure}[htb!]
    \centering
    \includegraphics[width=2.7in]{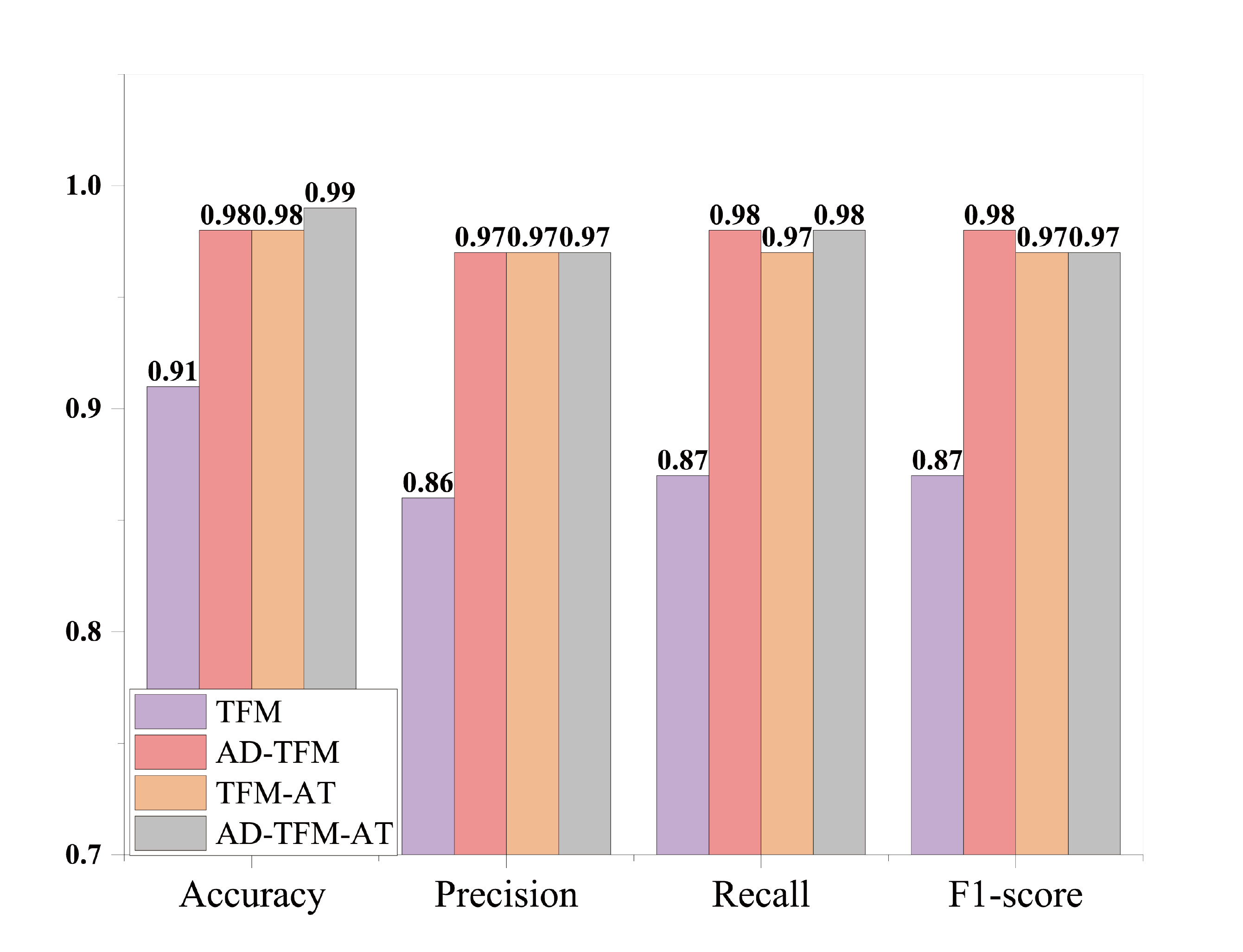}
    \caption{Ablation experiments results on SGAH dataset.}
    \label{GW XR metrics}
\end{figure}

\begin{figure*}[t!]
\centering
\subfloat[]{

\centering
\includegraphics[width=2.7in]{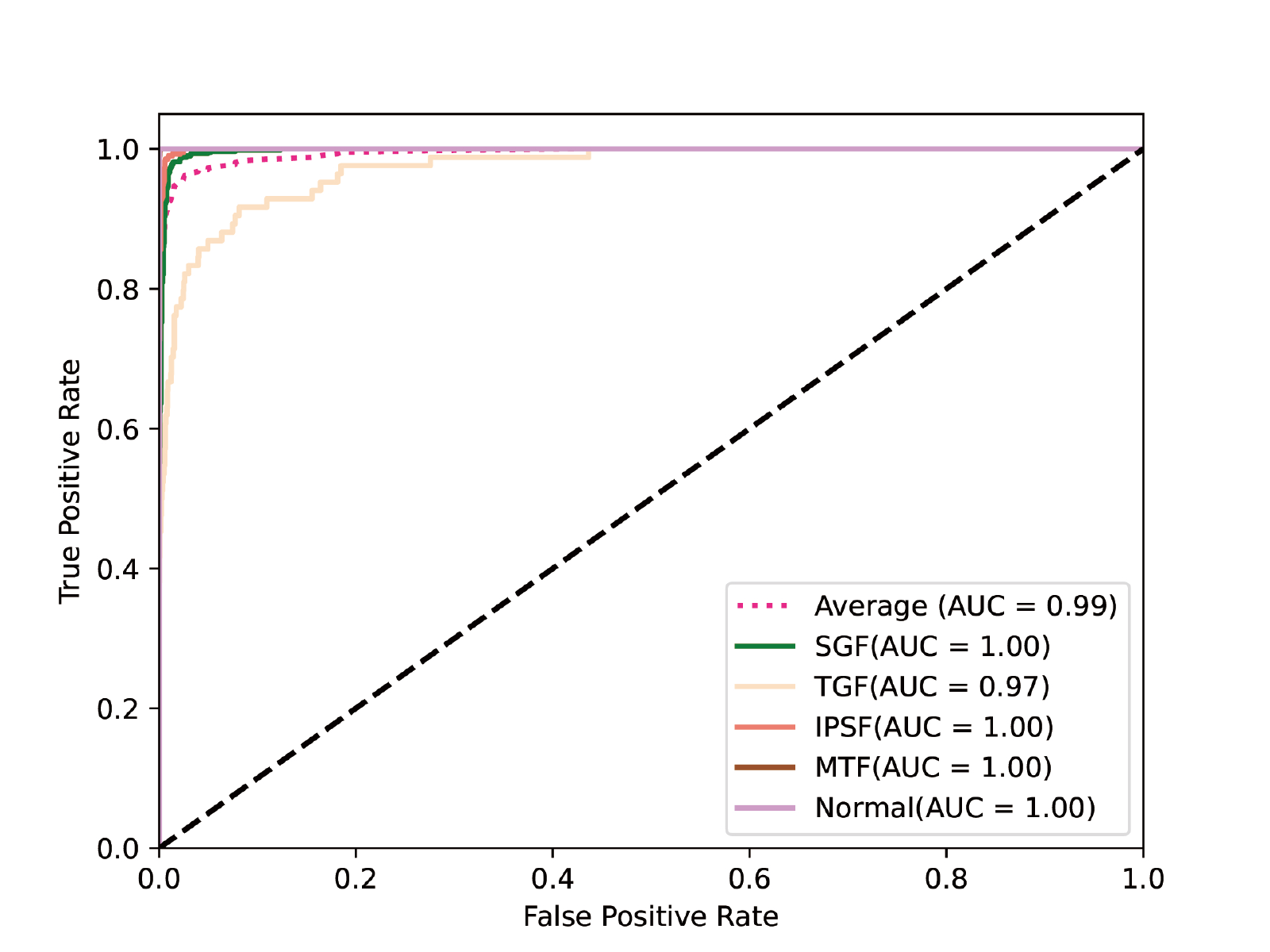}

}
\hspace{-8mm}
\subfloat[]{

\centering
\includegraphics[width=2.7in]{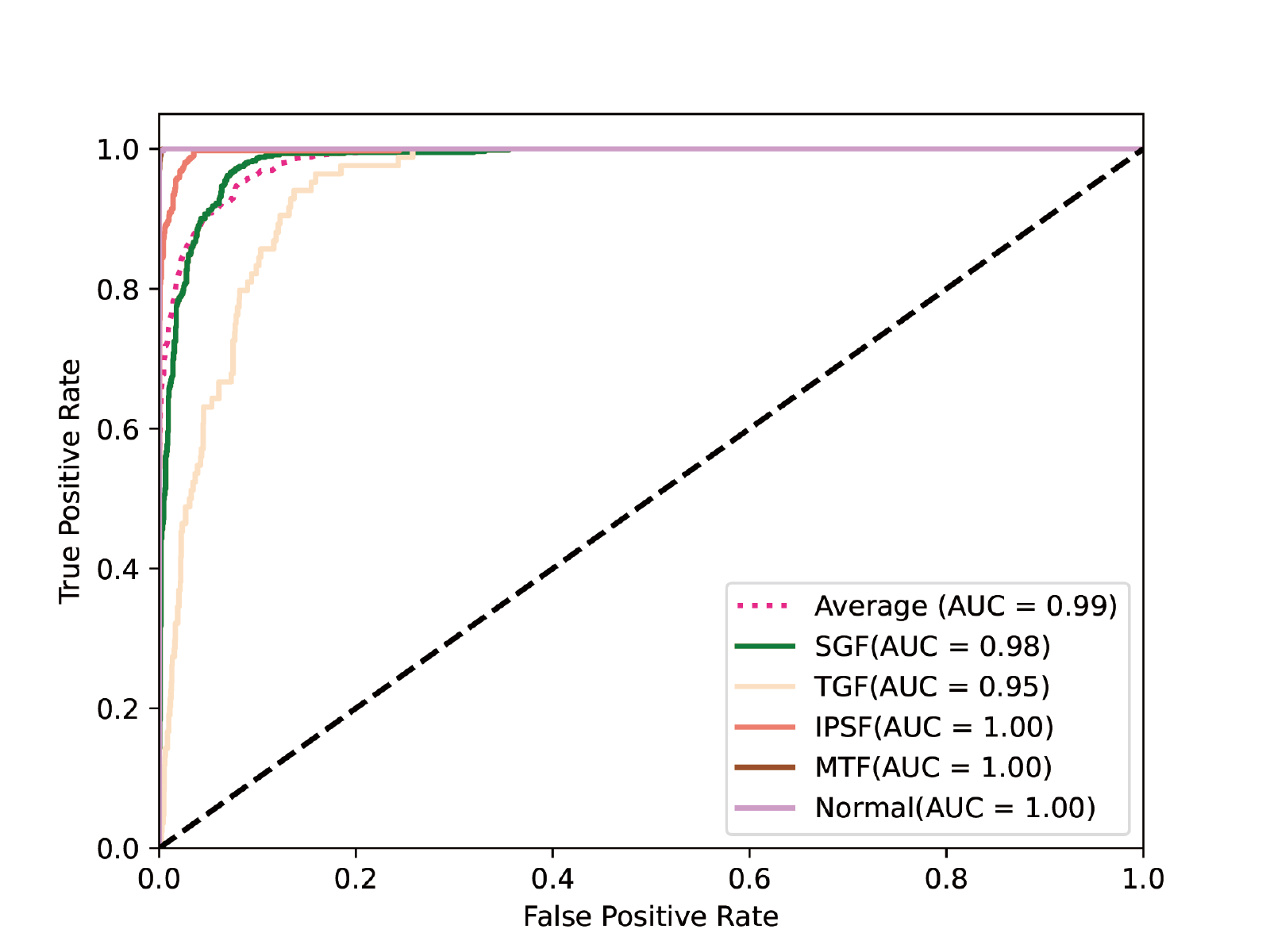}

}
\quad
\subfloat[]{

\centering
\includegraphics[width=2.7in]{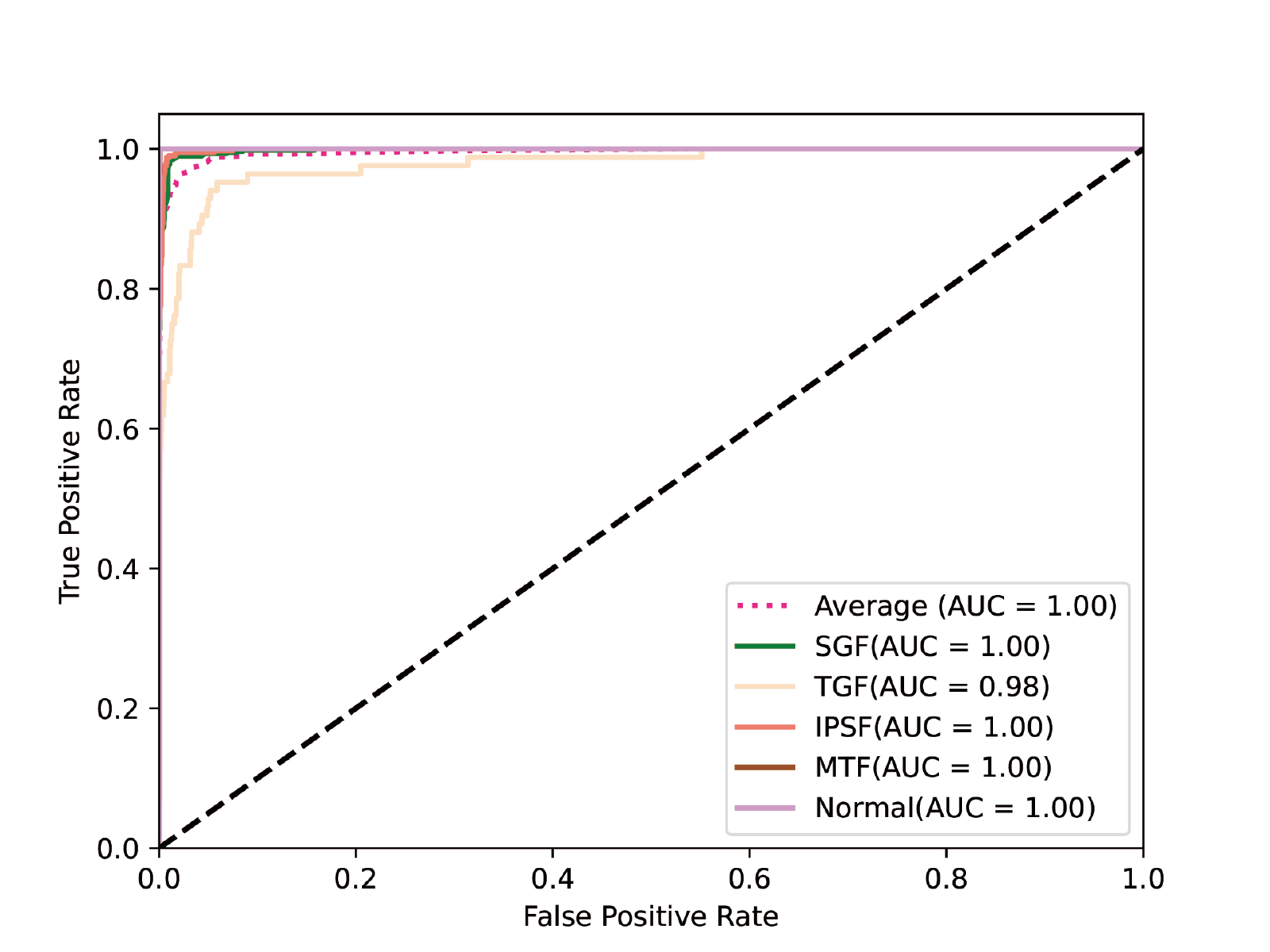}

}
\hspace{-8mm}
\subfloat[]{

\centering
\includegraphics[width=2.7in]{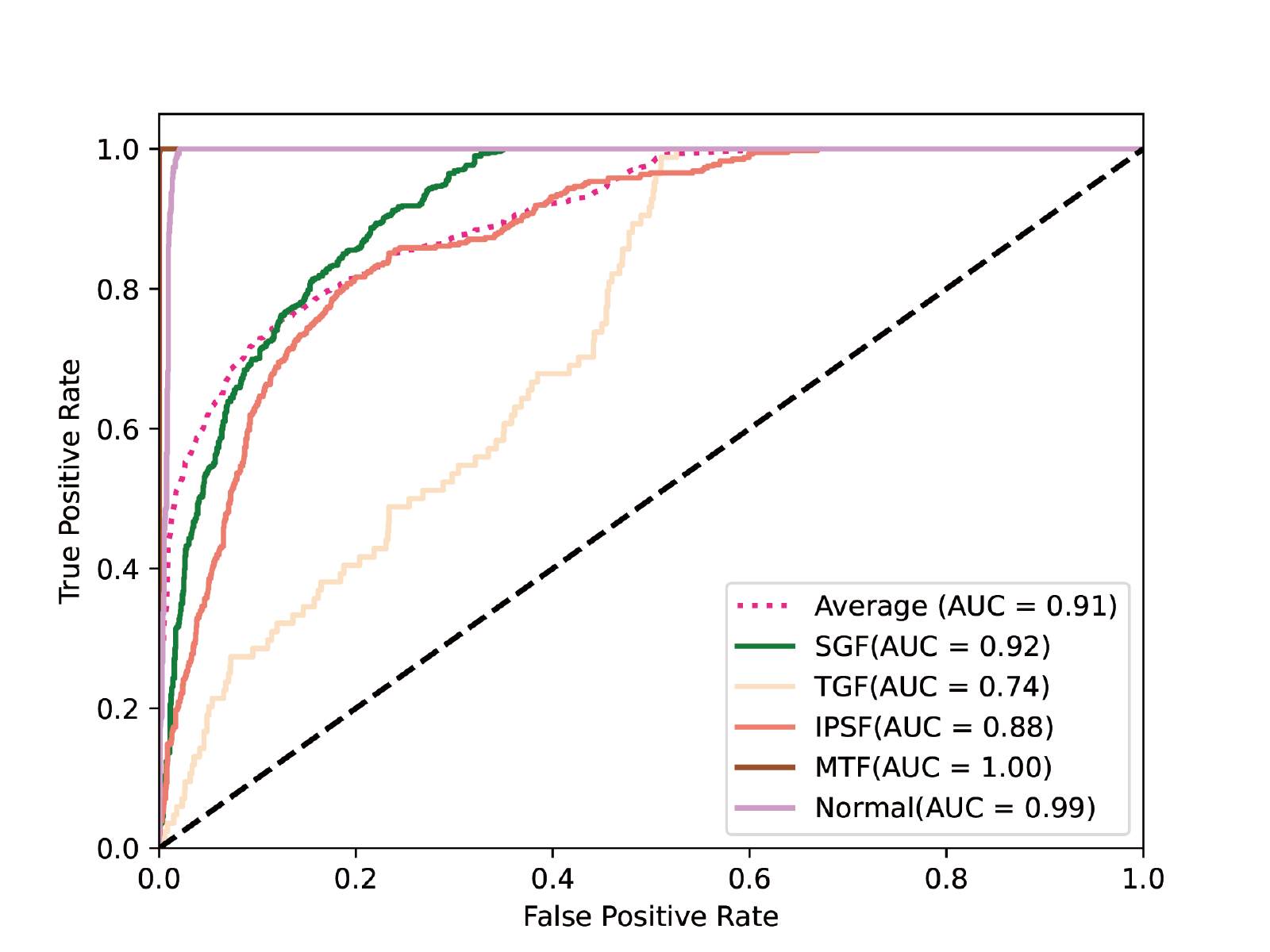}

}
\centering
\caption{ROC of ablation models on SGAH dataset. (a) ROC of AD-TFM-AT. (b) ROC of AD-TFM. (c) ROC of TFM-AT. (d) ROC of TFM.}
\label{ROC of GW XR}
\end{figure*}

\begin{figure}[htb!]
    \centering
    \includegraphics[width=3in]{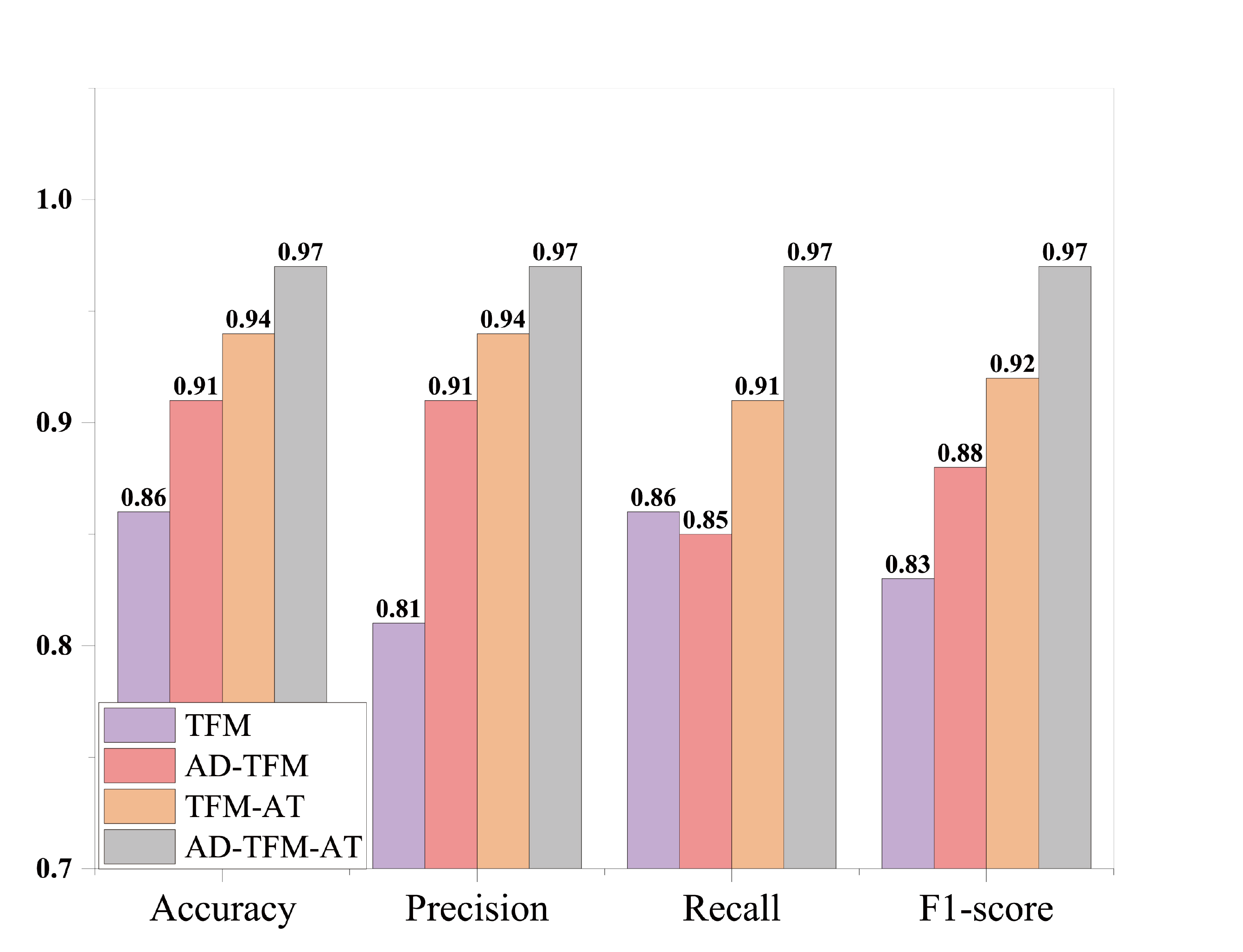}
    \caption{Ablation experiments results on IFPD dataset.}
    \label{WL XR metrics}
\end{figure}

\begin{figure*}[t!]
\centering
\subfloat[]{

\centering
\includegraphics[width=2.7in]{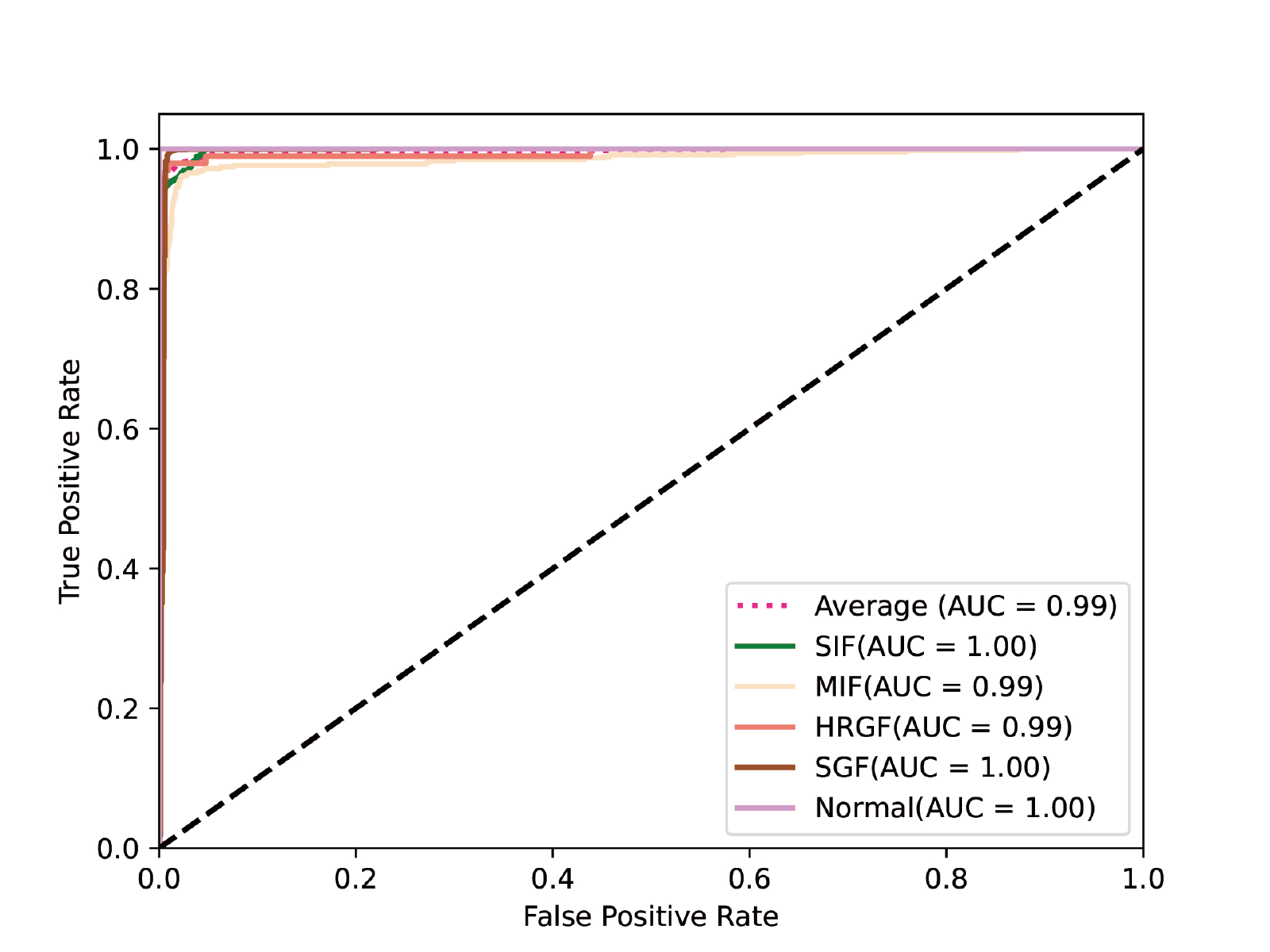}

}
\hspace{-8mm}
\subfloat[]{

\centering
\includegraphics[width=2.7in]{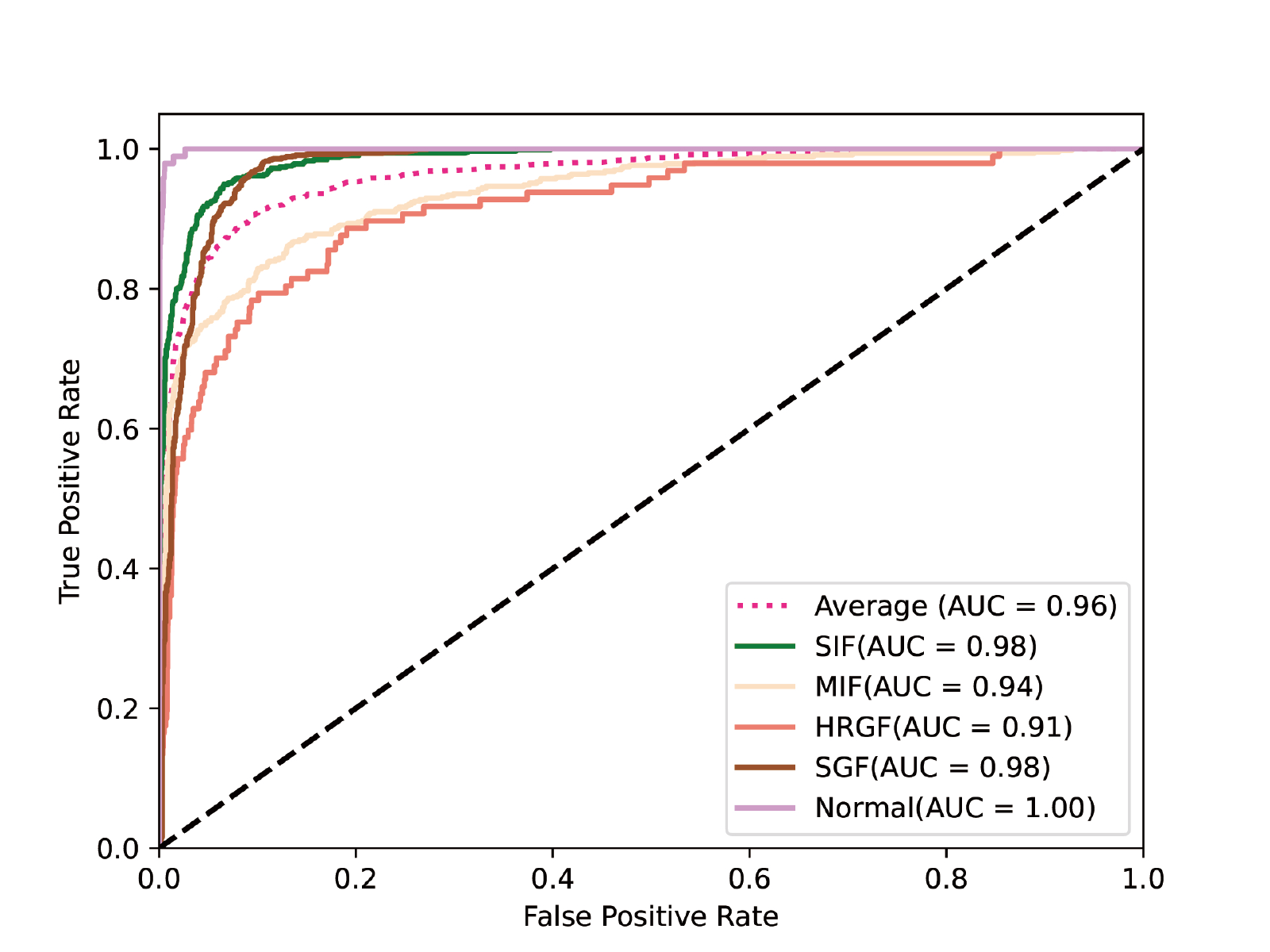}

}
\quad
\subfloat[]{

\centering
\includegraphics[width=2.7in]{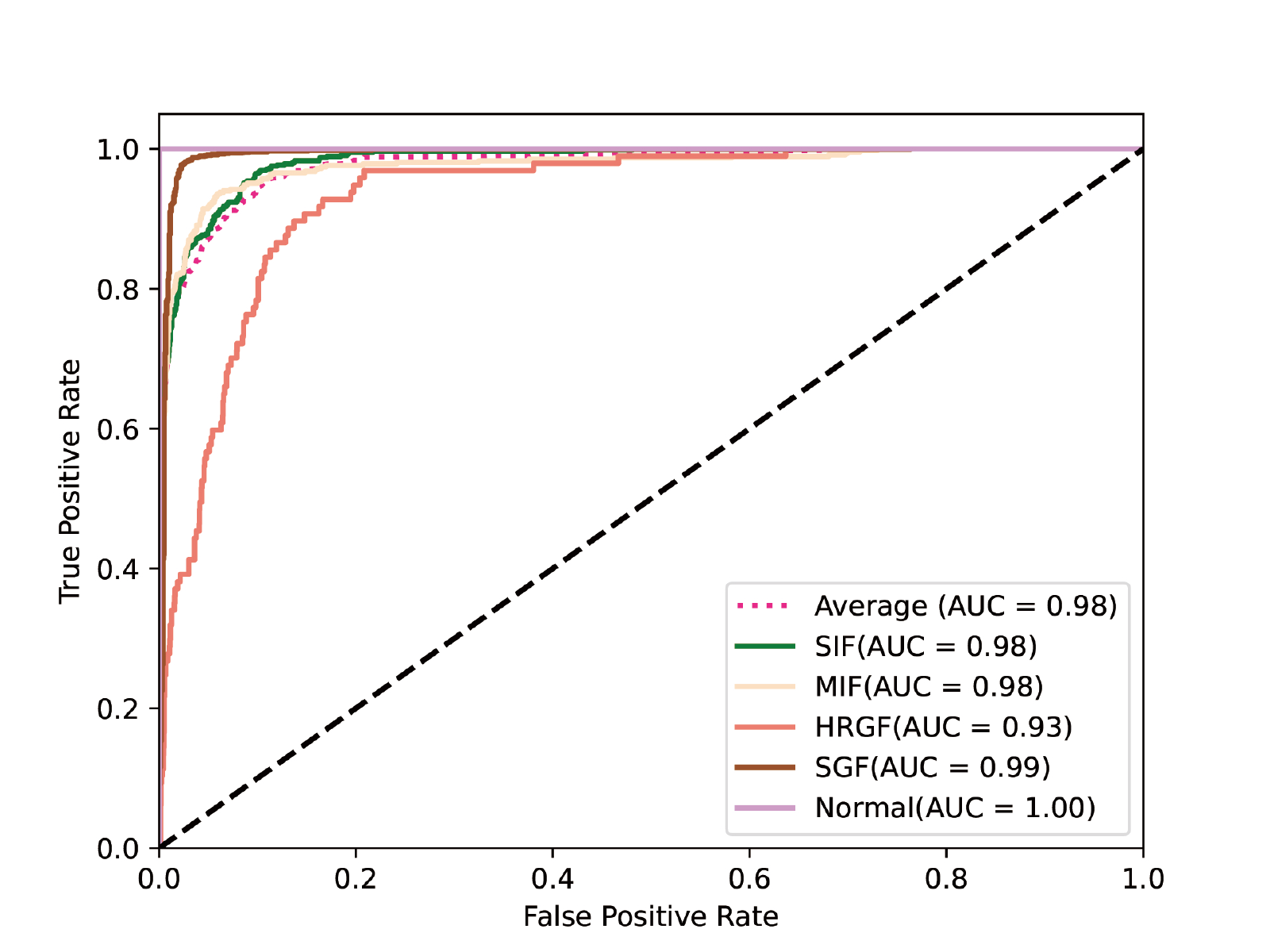}

}
\hspace{-8mm}
\subfloat[]{

\centering
\includegraphics[width=2.7in]{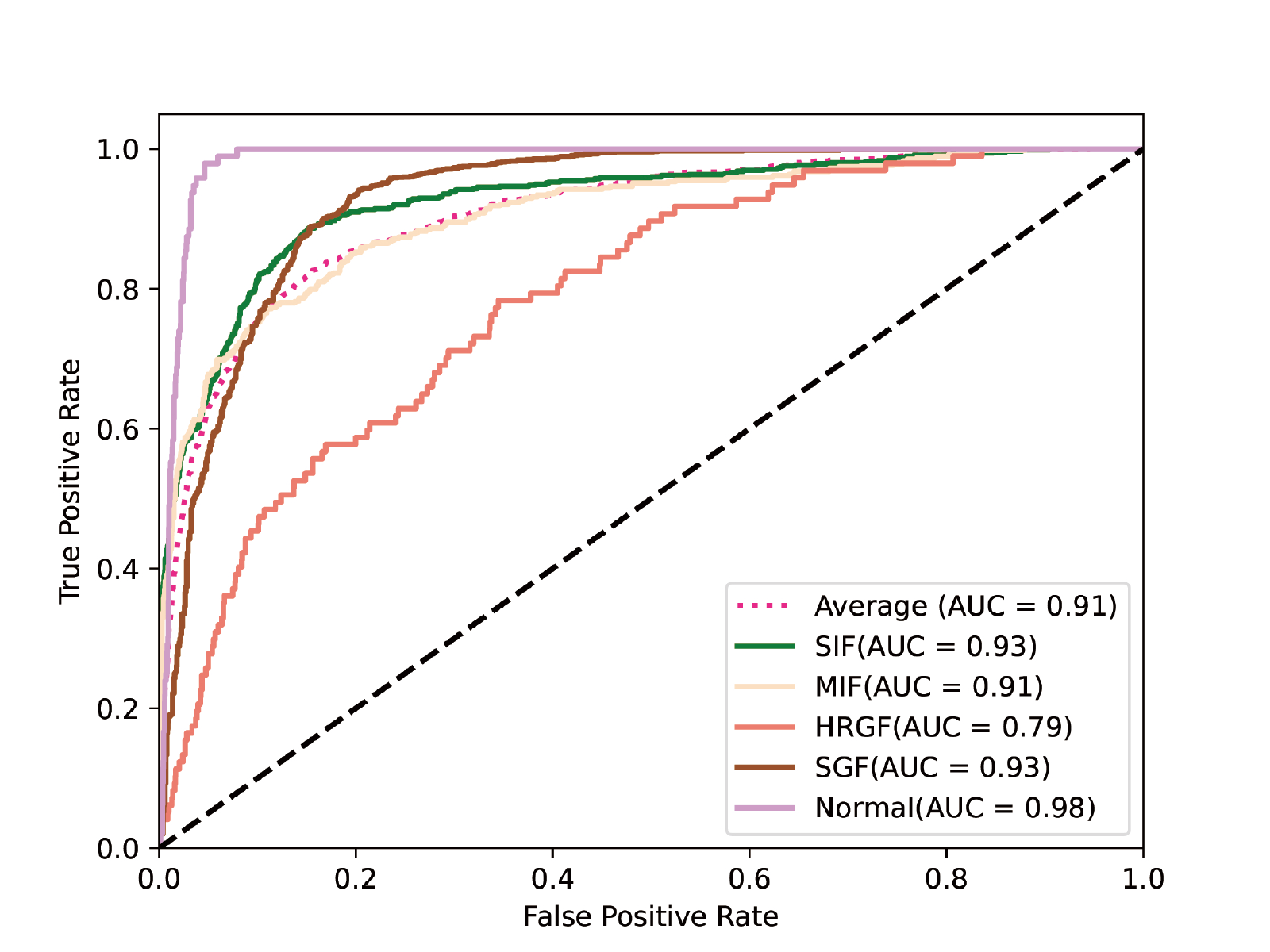}

}
\centering
\caption{ROC of ablation models on IFPD dataset. (a) ROC of AD-TFM-AT. (b) ROC of AD-TFM. (c) ROC of TFM-AT. (d) ROC of TFM.}
\label{ROC of WL XR}
\end{figure*}

\textbf{Ablation experiments results:}
We train and test the four models of TFM, AD-TFM, TFM-AT, AD-TFM-AT on the augmented IFPD and SGAH dataset, and the performance is shown in Fig. \ref{GW XR metrics} - Fig. \ref{ROC of WL XR}.

From Fig. \ref{GW XR metrics}, we can see that the model with the adaptive wavelet transform and the Attention mechanism performs better on SGAH dataset. From Fig. \ref{ROC of GW XR}, the Are Under Curve (AUC) of the TFM model for TGF is only 0.74, indicating the poor ability of the TFM model to detect this kind of fault. With the combination of adaptive wavelet transform and attention mechanism, the AUC of the TPF exceeds 0.90.
Both TGF and SGF belong to the ground fault, in which the three-phase voltage and current at the time of fault occurrence have similar characteristics, both of which show a voltage drop in the fault phase and distortion in the three-phase current. The AD-TFM and AD-TFM-AT models with the addition of adaptive wavelet transform have higher resolution than the three-phase voltage and current features extracted by the fixed parameter wavelet base in TFM. On the other hand, the TGF and SPGF fault durations are different, and the TFM-AT and AD-TFM-AT models with the added Attention mechanism focus on different time periods of fault information that the TFM lacks attention.

From Fig. \ref{WL XR metrics}, we can see that the AD-TFM-AT model with the introduction of adaptive wavelet transform and attention mechanism has significantly improved in terms of precision, accuracy, recall and F1 score when tested on IFPD data.
 
Among them, the four evaluation metrics of the AD-TFM-AT model are 0.1 higher than these of TFM model. From the ROC in Fig. \ref{ROC of WL XR}, we can see that the AUC under ROC curve for AD-TFM, TFM-AT and TFM models for HRGF is relatively low compared to other faults, while that of AD-TFM-AT is as high as 0.97. Therefore, the proposed method by combining the adaptive wavelet transform and attention mechanism increases the depth of the network and improves the accuracy and generalization ability of the model for incipient fault detection.

The above results show that adding adaptive wavelet transform to extract fault features at different times and frequencies can well deal with the non-stationary characteristics of incipient faults such as TGF, and the characterization of the fault feature vector for key fault information is enhanced by the attention mechanism, 
thus enabling the proposed AD-TFM-AT model to achieve high accuracy fault identification.

\textbf{Comparison with existing methods:}
The results of the comparison with existing methods are shown in Table~\ref{Comparison result}.

\begin{table*}[htb!]
\captionsetup{justification=centering}
\renewcommand\arraystretch{1}
 \centering
 \caption{Comparison results with existing models.}
 \label{Comparison result}
 \centering
    \setlength{\tabcolsep}{1mm}{
 \begin{tabular}{c cccc|cccc}
\toprule
\toprule
\multirow{2}{*}{Models}&
\multicolumn{4}{c}{SGAH Data}&\multicolumn{4}{c}{IFPD Data}\cr\cline{2-9}&Accuracy &Precision &Recall &F1score&Accuracy &Precision &Recall &F1score \\
\hline
AD-TFM-AT & \textbf{0.99} & \textbf{0.97} & \textbf{0.98} & \textbf{0.97}& \textbf{0.97} & \textbf{0.97} & \textbf{0.97} & \textbf{0.97} \\ 
SLI-CNN\cite{9702755}& 0.96 & 0.77 & 0.80 & 0.78& 0.93 & 0.93 & 0.82 & 0.84 \\ 
HLCL\cite{9094224} & 0.96 & 0.93 & 0.93 & 0.93 & 0.96 & 0.93 & 0.95 & 0.94 \\
Minirocket\cite{10.1145/3447548.3467231}& 0.96 & 0.81 & 0.97 & 0.80 & 0.91 & 0.84 & 0.93 & 0.88  \\
LSTM & 0.91 & 0.88 & 0.91 & 0.90& 0.93 & 0.90 & 0.93 & 0.91 \\
SVM  & 0.82 & 0.81 & 0.77 & 0.79& 0.85 & 0.83 & 0.77 & 0.80 \\
\bottomrule
    \end{tabular}}
    
\end{table*} 
From Table~\ref{Comparison result}, we can see that our proposed AD-TFM-AT model has the highest metrics on both SGAH and IFPD datasets, especially the accuracy reaches 0.99 and 0.97, respectively. In addition, on the SGAH dataset, Minirocket, SLI-CNN, and HLCL all achieve 0.96 accuracies, and SVM only has 0.82 lowest accuracies. Besides, Minirocket also has a high Recall up to 0.97, and HLCL also performs well. Note that SLI-CNN is only 0.77 on precision and F1-score 0.78.
On the IFPD dataset, the accuracy of the proposed AD-TFM-AT model reaches 0.97, which is the highest among all models. In addition, Minirocket's and LSI-CNN's accuracy reach 0.91 and 0.93, respectively. HLCL's accuracy and LSTM's accuracy reach 0.96 and 0.93, respectively. SVM's accuracy is 0.85, which is the lowest. 
This is because both TGF and SGF are one kind of ground fault, their three-phase voltage and three-phase current waveform have similar characteristics. The voltage waveform shows a drop in two phases. And another phase voltage maintains a normal state. Therefore, the waveform features of these two faults obtained by using convolution through images have similarity, which leads to low final classification accuracy and F1-score.

The above results show that among the existing fault detection methods, AD-TFM-AT has best performance. Instead, the Minirocket method and the SLI-CNN method use convolution to extract the features of time series, which lacks the analysis for non-stationary. Besides, they extract fault information accounts for a small component of the overall information, which may make wrong decisions. Therefore, it is not advisable to directly apply existing classification methods for fault classification. The AD-TFM-AT performs adaptive wavelet transform on the fault waveform data compared to other methods to achieve analysis of non-stationary. In addition, AD-TFM-AT also uses the attention mechanism to focus global information on fault. These make AD-TFM-AT the best performer on both datasets.

\section{Conclusion}
\label{sec_conclusion}

In this paper, we focus on incipient fault detection in power distribution systems and analyzed the non-stationary characteristics of incipient faults.
We propose an AD-TFM cell by embedding wavelet transform into the LSTM, to extract features in time and frequency domain from the non-stationary incipient fault signals. We make scale parameters and translation parameters of wavelet transform learnable to follow the dynamic input signals to analyse incipient fault with multi-resolution and multi-dimension analysis.
Based on the stacked AD-TFM cells, we design a AD-TFM-AT model to obtain more efficient fault features. In addition, we propose two data augmentation methods, namely phase switching and temporal sliding, to effectively enlarge the training datasets. Experimental results on two open datasets show that our proposed AD-TFM-AT model and data augmentation methods achieve better performance of incipient fault detection in power distribution system.

\section*{Acknowledgment}
This work is supported in part by grants from the National Natural Science Foundation of China (52077049, 51877060, 62173120), the Anhui Provincial Natural Science Foundation (2008085UD04, 2108085UD07, 2108085UD11), the 111 Project (BP0719039).

\ifCLASSOPTIONcaptionsoff
  \newpage
\fi

\bibliographystyle{IEEEtran}
\bibliography{incipient_fault}
\end{document}